\documentclass[12pt, draftclsnofoot, onecolumn]{IEEEtran}
\usepackage{color}
\usepackage{amsmath}
\usepackage{amsthm}
\usepackage{amssymb}
\usepackage{graphicx}
\usepackage{algorithm}
\usepackage{algorithmic}
\usepackage{cite}
\usepackage{setspace}
\newtheorem{theorem}{Theorem}

\newtheorem{remark}{Remark}

\ifCLASSINFOpdf
\else
\fi

\begin{document}
	\title{On the Doppler Squint Effect in OTFS Systems over Doubly-Dispersive Channels: Modeling and Evaluation}
	
	\author{
		Xuehan~Wang,~\IEEEmembership{Student~Member,~IEEE},
		Xu~Shi,~\IEEEmembership{Student~Member,~IEEE},
		Jintao~Wang,~\IEEEmembership{Senior~Member,~IEEE},
		and
		Jian~Song,~\IEEEmembership{Fellow,~IEEE}
		\thanks{Xuehan Wang, Xu Shi, Jintao Wang and Jian Song are with the Department of Electronic Engineering, Tsinghua University, Beijing, 100084, China (e-mail: wang-xh21@mails.tsinghua.edu.cn; shi-x19@mails.tsinghua.edu.cn; wangjintao@tsinghua.edu.cn; jsong@tsinghua.edu.cn).\par 
			This work was supported in part by Tsinghua University-China Mobile Research Institute Joint Innovation Center.\par 
		}
	}
	
	\maketitle
	\begin{spacing}{1.36}
	\begin{abstract} 
		Extensive work has demonstrated the excellent performance of orthogonal time frequency space (OTFS) modulation in high-mobility scenarios. Time-variant wideband channel estimation serves as one of the key compositions of OTFS receivers since the data detection requires accurate channel state information (CSI). In practical wideband OTFS systems, the Doppler shift brought by the high mobility is frequency-dependent, which is referred to as the Doppler Squint Effect (DSE). Unfortunately, DSE was ignored in overall prior estimation schemes employed in OTFS systems, which leads to severe performance loss in channel estimation and the consequent data detection. In this paper, we investigate DSE of wideband time-variant channel in delay-Doppler domain and concentrate on the characterization of OTFS channel coefficients considering DSE. The formulation and evaluation of OTFS input-output relationship are provided for both ideal and rectangular waveforms considering DSE. The channel estimation is therefore formulated as a sparse signal recovery problem and an orthogonal matching pursuit (OMP)-based scheme is adopted to solve it. Simulation results confirm the significance of DSE and the performance superiority compared with traditional channel estimation approaches ignoring DSE.
	\end{abstract}
	
	\IEEEpeerreviewmaketitle
	\section{Introduction}
	\label{sec:intro}
	Support of the ultra-reliable data transmission in high-mobility scenarios is required in the emerging fifth-generation (5G) and future wireless communication networks\cite{background_speed_ref,background_5g_nr_v2x,survey_require_data_mobility}, where the relative mobile velocity can be up to $500$ km/h \cite{raliway_speed_ref} for high-speed trains and $300$ km/h \cite{vechicle_speed_ref} for vehicles. Unfortunately, though orthogonal
	frequency division multiplexing (OFDM) technology has played a significant role in the mobile wireless communication systems to combat the inter-symbol interference (ISI) caused by multipath delay spread, it is non-trivial to mitigate the inter-carrier interference (ICI) brought by the high Doppler spread, which leads to severe performance degradation \cite{OFDM_performance_loss_ref} in realistic OFDM systems.\par 
	In order to improve the robustness of current communication schemes in high-mobility circumstances, orthogonal time frequency space (OTFS) modulation has been proposed in \cite{OTFS_propose} and drawn much concentration due to the significant advantages in linear time-variant (LTV) channels. In OTFS systems, the time-variant multipath channel is transformed into a time-invariant one in delay-Doppler domain and near-constant channel gain is provided consequently. Rather than modulating the data symbols in the time-frequency plane, the effective information symbols are processed over the delay-Doppler domain and experience full diversity over time and frequency, which makes data transmission more practical and reliable for both coded and uncoded scenarios\cite{OTFS_performance1,OTFS_performance3,otfs_signal_model,OTFS_performance2}. Considering both the ideal bi-orthogonal and the rectangular pulse-shape, the authors in \cite{otfs_signal_model} derived the input-output relationship and designed a message-passing-based receiver taking the sparse delay-Doppler channel coefficients into account. More efficient and reliable data detection schemes were investigated in \cite{low_complex_data_detection,low_complex_data_detection2,OTFS_frac_R1,OTFS_ZPdetect_reviewer1,OTFS_iterdetection_reviewer3,OTFS_crossdomain_reviewer4} based on the input-output analysis in \cite{otfs_signal_model}. Meanwhile, similar to the evolution of OFDM technology, a new path division multiple access for OTFS systems was proposed in \cite{OTFS_access} and index modulation in delay-Doppler domain has also been investigated in \cite{OTFS_IM1,OTFS_IM2}, which provided a new approach to promote the transmission efficiency and mitigate the peak-to-average power ratio further.\par 
	Acquisition of time-variant wideband channel state information (CSI) is critical for OTFS systems to detect the data symbols successfully, which has attracted plenty of attention\cite{OTFS_CE_PNpilot1_tf,OTFS_CE_PNpilot2_tf,OTFS_MPCE_reviewer2,OTFSmodel_cross,OTFS_CE_EMVB,OTFS_CE_VBI2,OTFS_CE_SBL,OTFS_CE_SBL2,OTFS_CE_impulse1,OTFS_CE_uplink1_OMP,OTFS_CE_threshold_classical,OTFS_CE_massivemimo_3domp,OTFS_CE_fractional,OTFS_CE_Windowdesign,OTFS_CE_subspace,OTFS_CE_dataaided1,OTFS_CE_dataaided2}. In general, the parameter estimation-based methods \cite{OTFS_MPCE_reviewer2,OTFSmodel_cross,OTFS_CE_EMVB,OTFS_CE_VBI2,OTFS_CE_SBL} outperform the direct channel coefficients estimation \cite{OTFS_CE_SBL2,OTFS_CE_impulse1,OTFS_CE_uplink1_OMP,OTFS_CE_threshold_classical,OTFS_CE_massivemimo_3domp,OTFS_CE_fractional}, where various off-grid estimation schemes such as the sparse Bayesian learning \cite{OTFS_CE_SBL} and message-passing algorithm\cite{OTFS_MPCE_reviewer2} have been demonstrated to be efficient in delay and Doppler extraction. Meanwhile, the impact of time-frequency window design was analyzed and Dolph-Chebyshev window was utilized to enhance the sparsity in delay-Doppler channel coefficients\cite{OTFS_CE_Windowdesign}. A set of transform-domain basis functions was introduced in \cite{OTFS_CE_subspace} to span a low-dimensional subspace for modeling the scattering-abundant channel, where the continuous Doppler spread channel (CDSC) such as the U-shaped one is taken into account.\par
	Most of the existing works on OTFS CSI acquisition are based on the conventional sparse multipath channel model in delay-Doppler domain \cite{otfs_signal_model}. However, as indicated in \cite{DSE_air}, there exists non-negligible Doppler difference across the subcarriers in wideband systems, which is referred to as the Doppler Squint Effect (DSE). Almost none of the existing work in OTFS channel estimation takes DSE into consideration since the basic input-output relationship employed are based on the analyses in \cite{otfs_signal_model,OTFSmodel_cross} where DSE is ignored. Since OTFS technology requires the assistance of 2D time-frequency modulation, the Doppler shift brought by high-mobility is subcarrier-dependent and the frequency-dependent phase offset caused by DSE will be accumulated through the time duration within one OTFS symbol, which causes severe performance degradation if channel estimation schemes ignoring DSE are deployed in practical OTFS systems. Classical investigations commonly ignore DSE by assuming that the ratio between the subcarrier spacing and the carrier frequency is small enough to ensure the near-constant Doppler shift on each subcarrier. Unfortunately, the significance of DSE in OTFS systems is approximately independent of this ratio, which is theoretically analyzed in this paper and forces us to reconsider the channel estimation in OTFS systems. \par
	To acquire more accurate CSI with DSE, we focus on the impact of DSE on delay-Doppler domain channel representation and OTFS input-output relationship, where both the ideal and the rectangular pulses are analyzed. An orthogonal matching pursuit (OMP)-based scheme is therefore adopted for more accurate wideband time-variant channel estimation. The contributions of this paper can be summarized as follows:
	\begin{itemize}
		\item \textbf{Channel response modeling:} Based on the input-output relationship of continuous LTV channels, we formulate the channel response in both the time-frequency domain and the delay-Doppler domain. DSE brings the time-frequency coupling which easily damages the widely-accepted assumption of sparsity in delay-Doppler domain. We prove that the single-path delay-Doppler response of wireless channel considering DSE is a constant-modulus waveform rather than the impulse supposed in traditional literature.
		\item \textbf{OTFS input-output formulation and evaluation:} Based on the delay-Doppler representation of wireless channel, the input-output relationship in OTFS systems considering DSE is derived, where both ideal bi-orthogonal and practical rectangular pulses are analyzed. For the ideal pulses, the time-frequency coefficients are formulated and an intuitive approximation is derived to evaluate the factors determining the significance of DSE, where we find the impact is only correlated with the discretized time-frequency size and the mobility speed. The extended delay and Doppler spread is then excavated by delay-Doppler domain input-output analysis. Meanwhile, the approximated formulation in delay-Doppler domain is attained with practical rectangular pulses adopted, where similar appearances can be found like the bi-orthogonal scenarios. It is worth pointing out that to the best of the authors’ knowledge, our work in this paper is the first one to consider DSE in OTFS systems.
		\item \textbf{The evaluation of DSE with channel estimation involved:} Combining the analysis of DSE with the widely-used impulse-based pilot technique, the target of channel estimation is formulated as a classic sparse signal recovery problem and OMP can be directly employed to estimate the multipath channel parameters. Simulation results demonstrate the significance of DSE in OTFS systems and the excellent performance of the estimation scheme considering DSE, which verifies the essentiality of our major contribution.
	\end{itemize} \par 
	The rest of the paper is organized as follows. The system model and general framework for OTFS signal analysis are presented in Section \ref{sec:system}, where the delay-Doppler representation of the multipath channel is formulated considering DSE. In Section \ref{sec:biopulses} and \ref{sec:recpulses}, we reformulate the OTFS input-output relationship with both the bi-orthogonal and the rectangular pulses, where the impact of DSE is taken into account. An OMP-based channel estimation scheme is adopted in Section \ref{sec:proposed} and the performance is evaluated by the simulations provided in Section \ref{sec:simulation}. At last, the conclusions are briefly drawn in Section \ref{sec:conclusion}. \par  
	\textit{Notations}: $\mathcal{A}$ is a set, $\mathbf{A}$ is a matrix, $\mathbf{a}$ is a column vector, $a$ is a scalar. $\mathbf{A}^H$ and $\mathbf{A}^{\dagger}$ denote its conjugate transposition and Moore-Penrose pseudo-inverse, respectively. $\mathbf{A}(i,j)$, $\mathbf{A}(i,:)$, $\mathbf{A}(:,j)$ are the $(i,j)$ component, $i^{th}$ row and $j^{th}$ column of $\mathbf{A}$. $||\mathbf{a}||_0$ and $||\mathbf{a}||_{2}$ denote the $l_0$-norm and $l_{2}$-norm of $\mathbf{a}$, respectively. $(\cdot)^{*}$ represents the conjugate operation while $(\cdot)_{N}$ denotes the modulus operation with respect to
	$N$. Finally, $\mathbb{I}_{\mathcal{A}}(x)$ is the indicator function for $x\in \mathcal{A}$.
	
	\section{System Model}
	\label{sec:system}
	\begin{figure*}
		\center{\includegraphics[width=0.75\linewidth]{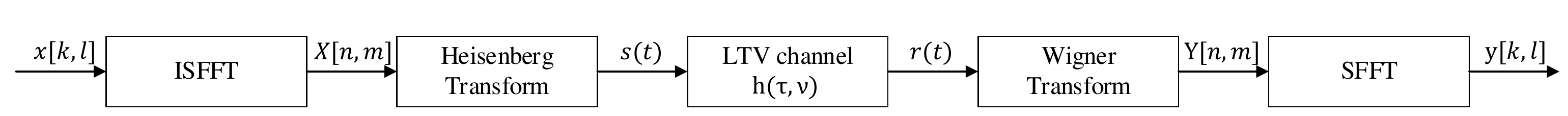}}
		\caption{OTFS transmitter and the receiver.}
		\label{Fig_system}	
	\end{figure*}
	In this section, we consider an OTFS system whose diagram is provided in Fig. \ref{Fig_system}. The impact of noise is disregarded to simplify the notation. Different from the analysis in \cite{otfs_signal_model}, the generalized delay-Doppler domain channel representation and the cross-ambiguity expression are modified. The model of multipath channel considering DSE is provided in this section as well, which serves as the basic of OTFS input-output analysis.   
	\subsection{OTFS Transmitter}
	\begin{itemize}
		\item[-] The time-frequency signal plane is discretized to a grid by sampling the time and frequency axes at intervals $T$ (seconds) and $\Delta f$ (Hz), respectively, i.e., $\Lambda=\{(nT,m\Delta f)|n=0,\cdots,N-1;m=0,\cdots,M-1\}$ for some integers $N,M>0$ and $T=\frac{1}{\Delta f}$.
		\item[-] The delay-Doppler information plane is discretized to a grid as $\Gamma=\{(\frac{k}{NT},\frac{l}{M\Delta f})|k=0,\cdots,N-1;l=0,\cdots,M-1\}$, where $\frac{1}{M\Delta f}$ and $\frac{1}{NT}$ represent the quantization resolution of the time delay and Doppler shift at the carrier frequency. 
	\end{itemize}\par
	Let $\tau_{\text{max}}$ denote the maximum delay spread and $\nu_{\text{max}}$ denote the maximum Doppler spread corresponding to the carrier frequency $f_c$, which can be denoted as
	\begin{equation}
		\scriptsize
		\tau_{\text{max}}=\frac{l_{\text{max}}}{M}T, \nu_{\text{max}}=\frac{k_{\text{max}}}{N}\Delta f.
	\end{equation}
	It is important to point out that $l_{\text{max}}$ and $k_{\text{max}}$ are only two notations, which can be not integers. To match the practical multipath wireless channel and exploit the potential of OTFS resolution, we require at least that $l_{\text{max}}>1$ and $k_{\text{max}}>1$. \par 
	Consider a sequence of $NM$ modulated data symbols $\{x[k,l]|k=0,\cdots,N-1;l=0,\cdots,M-1\}$, which are arranged on the delay-Doppler grid. 
	Each symbol $x[k,l]$ is from a modulation alphabet of size $Q$. The OTFS transmitter first maps the information symbols $x[k,l]$ on the delay-Doppler grid $\Gamma$ to $X[n,m]$ on the time-frequency grid $\Lambda$. The inverse symplectic finite Fourier transform (ISFFT) is employed in this process as follows:
	\begin{equation}
		\label{ISFFT}
		\scriptsize
		X[n,m]=\frac{1}{\sqrt{NM}}\sum_{k=0}^{N-1}\sum_{l=0}^{M-1}x[k,l]e^{j2\pi (\frac{nk}{N}-\frac{ml}{M})},
	\end{equation}
	for $n=0,\cdots,N-1,m=0,\cdots,M-1$.\par
	After that, the discretized time-frequency symbol sequence is transformed to a continuous baseband time waveform utilizing the Heisenberg transform as
	\begin{equation}
		\scriptsize
		s(t)=\sum_{n=0}^{N-1}\sum_{m=0}^{M-1}X[n,m]g_{\text{tx}}(t-nT)e^{j2\pi m\Delta f (t-nT)},
		\label{Heisenberg_transform}
	\end{equation}
	where $g_{\text{tx}}(t)$ denotes the transmit pulse shape.
	\subsection{Multipath Channel Model considering DSE}
	After the up-conversion, the continuous-time passband $\Re \big\{s(t)e^{j2\pi f_{c}t} \big\}$ is sent from the transmitter to the receiver via $N_{P}$ incident paths, where $f_{c}$ is the carrier frequency. The received passband signal can be derived by \cite{channel_tf_orig}
	\begin{equation}
		\scriptsize
		\widetilde{r}(t)=\Re \bigg\{\sum_{i=1}^{N_{P}}\widetilde{\beta_{i}}s(t-(\tau_{i}-\frac{v_{i}}{c}t))e^{j2\pi f_{c}(t-(\tau_{i}-\frac{v_{i}}{c}t))}\bigg\},
	\end{equation}
	where $\widetilde{\beta_{i}}$, $\tau_{i}$ and $v_{i}$ represent the attenuation, propagation delay and velocity with which the $i^{th}$ path length is decreasing. The total delay of the $i^{th}$ path can be represented as $\tau_{i}(t)=\tau_{i}-\frac{v_{i}}{c}t$. Let $\nu_{i}=\frac{v_{i}}{c}f_{c}$ denote the Doppler shift at the carrier frequency of the $i^{th}$ path and remove the carrier $e^{j2\pi f_{c}t}$, the baseband received signal is
	\begin{equation}
		\scriptsize
		r(t)=\sum_{i=1}^{N_{P}}\widetilde{\beta_{i}}e^{-j2\pi \tau_{i}f_{c}}e^{j2\pi\nu_{i}t}s(t-(\tau_{i}-\frac{\nu_{i}}{f_{c}}t)).
		\label{receive_signal_DSE_channel}
	\end{equation}\par
	We first consider to employ the time-varying frequency response $H(t,f)$ to characterize the multipath channel, where the received signal $r(t)$ can be derived \cite{channel_model_tf_cs} as
	\begin{equation}
		\scriptsize
		r(t)=\int H(t,f)S(f)e^{j2\pi tf}df
		\label{tf_receive},
	\end{equation}
	where $S(f)$ is the Fourier transform of $s(t)$. Combining \eqref{receive_signal_DSE_channel} and \eqref{tf_receive}, $H(t,f)$ can be derived as
	\begin{equation}
		\scriptsize
		H(t,f)=\sum_{i=1}^{N_{P}}
		\beta_{i}
		e^{j2\pi\frac{\nu_{i}}{f_{c}}(f_{c}+f)t} e^{-j2\pi f\tau_{i}},
		\label{channel_response_tf_DSE}
	\end{equation}
	where $\beta_{i}=\widetilde{\beta_{i}}e^{-j2\pi f_{c}\tau_{i}}$. \eqref{channel_response_tf_DSE} indicates that the Doppler shift brought by high-mobility is subcarrier-dependent and the frequency-dependent phase offset caused by DSE will be accumulated through the time duration within one OTFS symbol, which is referred to as the Doppler squint effect. On the other hand, DSE brings the time-frequency coupling, which easily destroys the sparse representation in delay-Doppler domain. It is disastrous to the existing channel estimation and data detection schemes since the sparse representation of delay-Doppler channel is usually treated as a basic assumption in prior work. The following analysis of the baseband delay-Doppler response of the wireless multipath channel will help clarify this declaration.\par 
	Since the delay-Doppler channel response $h(\tau,\nu)$ is the 2D symplectic Fourier transform (SFT) of $H(t,f)$, the received signal can also be represented \cite{channel_model_taomiu_classic} as
	\begin{equation}
		\scriptsize
		r(t)=\iint\ h(\tau,\nu)s(t-\tau)e^{j2\pi \nu t}d\tau d\nu
		\label{dd_receive}.
	\end{equation} \par
	Compared with the delay-Doppler representation of LTV channel in \cite{otfs_signal_model}, the modification is embodied in the phase shift caused by the Doppler shift, where $e^{j2\pi \nu (t-\tau)}$ is replaced with $e^{j2\pi \nu t}$. The modified phase shift can be easily found in \eqref{receive_signal_DSE_channel}, in which the phase shift due to the Doppler shift is $e^{j2\pi\nu_{i}t}$ rather than $e^{j2\pi\nu_{i}(t-\tau)}$. The delay-Doppler representation in \eqref{dd_receive} is same as the analysis in \cite{channel_model_taomiu_classic,OTFSmodel_cross,OTFS_tfpilot_ddresponseprove}. $h(\tau,\nu)$ of the multipath channel considering DSE is formulated as the following theorem.\par
	\begin{theorem}
		\label{th1_dd_channel}
		\rm{The delay-Doppler response of LTV channel can be represented as}
		\begin{equation}
			\scriptsize
			h(\tau,\nu)=\sum_{i=1}^{N_{P}}h_{i}(\tau,\nu),
			\label{repre_dd_channel}
		\end{equation}
		where $h_{i}(\tau,\nu)$ denotes the delay-Doppler response of each path as
		\begin{equation}
			\scriptsize
			h_{i}(\tau,\nu)=
			\begin{cases}
				\beta_{i}\delta(\tau-\tau_{i})\delta(\nu),&\nu_{i}=0\\
				\beta_{i}\Big|\frac{f_{c}}{\nu_{i}}\Big|e^{j2\pi\frac{f_{c}}{\nu_{i}}(\tau-\tau_{i})(\nu-\nu_{i})},&\nu_{i}\neq 0
			\end{cases}.
			\label{h_DD_wirelesschannel}
		\end{equation}
		\begin{IEEEproof}
			The proof is finished in Appendix \ref{th2_channel_dd_proof}.
		\end{IEEEproof}
	\end{theorem} \par 
	Through Theorem \ref{th1_dd_channel}, it is obvious that the single-path delay-Doppler response is a constant-modulus waveform rather than an impulse in traditional literature when $\nu_{i}\neq 0$. It brings significant difference in OTFS input-output relationship even though ideal pulse shape can be achieved, which will be illustrated in detail in Section \ref{sec:biopulses} and \ref{sec:recpulses}. Meanwhile, the notation $\tau_{i}=\frac{l_{i}}{M}T$ and $\nu_{i}=\frac{k_{i}}{N}\Delta f$ are adopted
	to simplify the representation in the following part, where $l_{i}$ and $k_{i}$ are not necessarily integers in common analysis.
	\begin{remark}
		If the frame duration is small or the mobility is slow enough, then $\tau_{i}-\frac{\nu_{i}}{f_{c}}t\approx \tau_{i}$ will hold true within a frame duration $T_{s}$, i.e., the phase modification caused by frequency-dependent Doppler shift $e^{j2\pi\frac{\nu_{i}}{f_{c}}ft}$ within a frame can be ignored, which will deduce the widely-employed model in OFDM and OTFS systems \cite{otfs_signal_model,OTFS_CE_Windowdesign,OTFS_CE_massivemimo_3domp} as $h(\tau,\nu)=\sum_{i=1}^{N_{P}}\beta_{i}\delta(\tau-\tau_{i})\delta(\nu-\nu_{i})$. In traditional OFDM systems, it is acceptable to ignore DSE because $\Delta f\ll f_{c}$ means the Doppler offset is negligible and the deviation will be reset in next OFDM symbol. However, DSE is significant in OTFS systems since the technique of 2D-modulation is employed, which means the deviation will be accumulated along the time axis. Taking the typical value $v=500$ km/h, $M=2048$ and $N=128$ as an example, DSE leads to a maximum phase offset about $e^{j2\pi\frac{\nu_{i}}{f_{c}}\times M\Delta f \times NT}=e^{j0.24\pi}$, which is non-negligible. 
	\end{remark}  
	\subsection{OTFS Receiver}
	Corresponding to the signal model in \eqref{dd_receive}, a matched filter computes the cross-ambiguity function $A_{g_{\text{rx}},r}$ as
	\begin{equation}
		\scriptsize
		Y(t,f)=A_{g_{\text{rx}},r}=\int g_{\text{rx}}^{*}(t^{\prime}-t)r(t^{\prime})e^{-j2\pi ft^{\prime}}dt^{\prime},
		\label{cross-ambiguity}
	\end{equation}  
	which is consistent with the cross-ambiguity module widely deployed in the Radar systems \cite{OTFSmodel_cross}.\par
	The output of matched filter $Y(t,f)$ is sampled as 
	\begin{equation}
		\scriptsize
		Y[n,m]=Y(t,f)|_{t=nT,f=m\Delta f},
		\label{tfsymbol_receiver}
	\end{equation} 
	for $0\leq n\leq N-1$ and $0\leq m\leq M-1$. The operations in \eqref{cross-ambiguity} and \eqref{tfsymbol_receiver} are referred to as Wigner transform. Combining the analysis in \cite{otfs_signal_model} and \cite{OTFSmodel_cross}, the time-frequency input-output relationship can be derived as $Y[n,m]=\sum_{n^{\prime}=0}^{N-1}\sum_{m^{\prime}=0}^{M-1}H_{n,m}[n^{\prime},m^{\prime}]X[n^{\prime},m^{\prime}]$, where we have 
	\begin{equation}
		\scriptsize
		\begin{aligned}
			H_{n,m}[n^{\prime},m^{\prime}]=\iint h(\tau,\nu)A_{g_{\text{rx}},g_{\text{tx}}}((n-n^{\prime})T-\tau,(m-m^{\prime})\Delta f-\nu)e^{j2\pi(\nu -m\Delta f)(\tau+n^{\prime}T)}d\tau d\nu.	
		\end{aligned}
		\label{H_tf_general}
	\end{equation}\par
	After the Wigner transform, the symplectic finite Fourier transform (SFFT) is executed as
	\begin{equation}
		\scriptsize
		y[k,l]=\frac{1}{\sqrt{NM}}\sum_{n=0}^{N-1}\sum_{m=0}^{M-1}Y[n,m]e^{-j2\pi (\frac{nk}{N}-\frac{ml}{M})}.
		\label{SFFT}
	\end{equation} 	
\section{OTFS With Bi-Orthogonal Waveforms}
	\label{sec:biopulses}
	In this section, we formulate the modified input-output relationship in OTFS system with bi-orthogonal waveform based on the analysis of delay-Doppler channel response considering DSE. The factors determining the significance of DSE will also be addressed. For ease of illustration, The impact of noise is disregarded to simplify the notation like Section \ref{sec:system}. We consider the OTFS system with $MN<10^{6}$, which is easily compatible with the existing wireless communication networks\cite{otfs_signal_model}, e.g., $M=512$ and $N=128$ with the carrier frequency $f_{c}=4$GHz and the subcarrier spacing $\Delta{f}=15$kHz. Let $p_{i}=\frac{f_{c}}{\nu_{i}}$ for ease of illustration, where we have $|p_{i}|\geq\frac{c}{v}>10^{6}$ because the relative velocity $v$ is less than $1000$ km/h in realistic scenarios.\par  
	If $A_{g_{\text{rx}},g_{\text{tx}}}(t,f)=\mathbb{I}_{ [-t_{\text{max}},t_{\text{max}}]}(t)\mathbb{I}_{[-f_{\text{max}},f_{\text{max}}]}(f)$, the pulses $g_{\text{rx}}(t)$ and $g_{\text{tx}}(t)$ are said to satisfy bi-orthogonal property.
	%\begin{equation}
	%	\scriptsize
	%		A_{g_{\text{rx}},g_{\text{tx}}}(t,f)=
	%	\begin{cases}
	%		1,&(t,f)\in [-t_{\text{max}},t_{\text{max}}]\times[-f_{\text{max}},f_{\text{max}}]\\
	%		0,&\text{elsewhere}\\
	%	\end{cases}.
	%\end{equation}\par 
	Unfortunately, bi-orthogonal pulses do not exist in practical scenarios according to the Heisenberg uncertainty principle. Nevertheless, it is essential to investigate the property of OTFS system with bi-orthogonal waveforms since it serves as a tight performance bound\cite{otfs_signal_model} for OTFS systems with practical waveforms such as the rectangular pulses. Meanwhile, the analysis framework is similar regardless of the waveforms employed.
	\subsection{Time-Frequency Domain Analysis}
	In this subsection, we focus on the difference in OTFS time-frequency input-output relationship brought by DSE. The foundation of this part has been proposed in Section \ref{sec:system}. Since the impact of each path can be analyzed separately similar to Appendix \ref{th2_channel_dd_proof}, we concentrate on the analysis of single-path response, which is referred to as $H_{n,m}[n^{\prime},m^{\prime}]=\sum_{i=1}^{N_{P}}H_{n,m}^{i}[n^{\prime},m^{\prime}]$. Besides, the deduction when $\nu_{i}=0$ is trivial and the inference for $\nu_{i}<0$ is similar to the scenario when $\nu_{i}>0$, which drives us to consider only $\nu_{i}>0$ in this part to simplify the notation.\par 
	Combining \eqref{H_tf_general} with \eqref{h_DD_wirelesschannel}, we first provide Theorem \ref{th2_H_tf_DSE_singlepath} as the basis of analysis framework.\par 
	\begin{theorem}
		\rm
		\label{th2_H_tf_DSE_singlepath}
		The time-frequency coefficients $H_{n,m}^{i}[n^{\prime},m^{\prime}]$ for OTFS system considering DSE can be derived as
		\begin{equation}
			\scriptsize
			\begin{aligned}
				H_{n,m}^{i}[n^{\prime},m^{\prime}]&=\beta_{i}p_{i}e^{j2\pi p_{i}\tau_{i}\big(\nu_{i}-(m-m^{\prime})\Delta f\big)}\int_{\tau_{1}^{\prime}}^{\tau_{2}^{\prime}}\frac{\sin\Big(2\pi f_{\text{max}}\big((1+p_{i})\tau-(p_{i}\tau_{i}-n^{\prime}T)\big)\Big)}{\pi \big((1+p_{i})\tau-(p_{i}\tau_{i}-n^{\prime}T)\big)}e^{-j2\pi\tau\big(f_c+m^{\prime}\Delta f-p_{i}(m-m^{\prime})\Delta f\big)}d\tau
			\end{aligned},
			\label{th2_eq_H_tf_DSE_singlepath}
		\end{equation}
		where we have $\tau_{1}^{\prime}=(n-n^{\prime})T-t_{\text{max}}$ and $\tau_{2}^{\prime}=(n-n^{\prime})T+t_{\text{max}}$ due to the finite-support property of the cross-ambiguity of bi-orthogonal pulse shape filter.
		\begin{IEEEproof}
			The proof is provided in Appendix \ref{th2_proof_apendix}.
		\end{IEEEproof}
	\end{theorem}\par 
	Theorem \ref{th2_H_tf_DSE_singlepath} informs that $H_{n,m}^{i}[n^{\prime},m^{\prime}]$ is the Fourier transformation of sinc function which is truncated to an finite interval. Assuming that only integer delay and Doppler are included in the wireless channel, we will demonstrate subsequently that if we choose $t_{\text{max}}$, $f_{\text{max}}$ and system parameters which satisfy the constraints as follows:
	\begin{equation}
		\scriptsize
		\begin{cases}
			\tau_{\text{max}}+\frac{T}{M}<t_{\text{max}}<\frac{T}{2}\\
			2\nu_{\text{max}}+\frac{\Delta f}{N}<f_{\text{max}}<\frac{\Delta f}{2}\\
		\end{cases}
		\label{system_constraints},
	\end{equation}
	$H_{n,m}^{i}[n^{\prime},m^{\prime}]\approx 0$ can be achieved when $n\neq n^{\prime}$ or $m\neq m^{\prime}$. At the meantime, the condition in \eqref{system_constraints} is near-optimal to some extent, which will be depicted in detail subsequently. The property of sinc function in \eqref{th2_eq_H_tf_DSE_singlepath} is explored by claiming two essential parameters. Let $x$ and $B$ denote the peak location and the mainlobe width to the nulls of sinc function in \eqref{th2_eq_H_tf_DSE_singlepath}, we have 
	\begin{equation}
		\scriptsize
		\begin{cases}
			x=\frac{p_{i}\tau_{i}-n^{\prime}T}{1+p_{i}}\\
			B=\frac{1}{f_{\text{max}}(1+p_{i})}.
		\end{cases}
		\label{property_sinc}
	\end{equation}\par 
	Let us begin with the discussion when $n\neq n^{\prime}$. If $n>n^{\prime}$, i.e., $n-n^{\prime}\geq 1$, we have
	\begin{equation}
		\scriptsize
		\begin{aligned}
			\tau_{1}^{\prime}-(x+\frac{B}{2})&=(n-n^{\prime})T-t_{\text{max}}-\frac{p_{i}\tau_{i}-n^{\prime}T}{1+p_{i}}-\frac{1}{2f_{\text{max}}(1+p_{i})}\\
			&\overset{(a)}{\geq} T-t_{\text{max}}-\tau_{i}-\frac{1}{2f_{\text{max}}(1+p_{i})}\\
			&\overset{(b)}{>}\frac{T}{M}\big(M-\frac{M}{2}-(\frac{M}{2}-1)-\frac{M}{2f_{\text{max}}(1+p_{i})T}\big)=\frac{T}{M}\big(1-\frac{1}{2(2k_{\text{max}}+1)}\big)>0\\
		\end{aligned}
		\label{ana_n>n'},
	\end{equation}
	where $(b)$ is obtained from \eqref{system_constraints} and $p_{i}>10^{6}>MN$, which leads the inequality as follows:
	\begin{equation}
		\scriptsize
		\begin{aligned}
			\frac{M}{2f_{\text{max}}(1+p_{i})T}<\frac{MN}{2(2k_{\text{max}}+1)(1+p_{i})}<\frac{1}{2(2k_{\text{max}}+1)}
		\end{aligned}.
		\label{ana_B}
	\end{equation}\par
	If $n<n^{\prime}$, i.e., $n-n^{\prime}\leq -1$, we have
	\begin{equation}
		\scriptsize
		\begin{aligned}
			(x-\frac{B}{2})-\tau_{2}^{\prime}&=\frac{p_{i}\tau_{i}-n^{\prime}T}{1+p_{i}}-\frac{1}{2f_{\text{max}}(1+p_{i})}-(n-n^{\prime})T-t_{\text{max}}\\
			&\geq T-t_{\text{max}}+\frac{p_{i}\tau_{i}}{1+p_{i}}-\frac{n^{\prime}T}{1+p_{i}}-\frac{1}{2f_{\text{max}}(1+p_{i})}\geq T-t_{\text{max}}-\frac{n^{\prime}T}{1+p_{i}}-\frac{1}{2f_{\text{max}}(1+p_{i})}\\
			&\overset{(a)}{>} \frac{T}{M}\big(M-\frac{M}{2}-1-\frac{1}{2(2k_{\text{max}}+1)}\big)\geq \frac{T}{M}(\frac{M}{2}-1-\frac{1}{2(2k_{\text{max}}+1)})>0		
		\end{aligned},
		\label{ana_n<n'}	
	\end{equation}
	where $(a)$ is obtained from $p_{i}>10^{6}>MN$ , $n^{\prime}<N$ and $t_{\text{max}}<\frac{T}{2}$.\par 
	From the analysis \eqref{ana_n>n'} and \eqref{ana_n<n'}, we acquire that the mainlobe never exists in the integral interval of \eqref{th2_eq_H_tf_DSE_singlepath} if $n\neq n^{\prime}$. Moreover, we can derive that the minimum distance between the integral interval and the mainlobe $d$ has the following property as:
	\begin{equation}
		\scriptsize
		\begin{aligned}
			\frac{d}{B/2}> \frac{2T}{MB}\big(1-\frac{1}{2(1+2k_{\text{max}})}\big)>\frac{1+p_{i}}{MN}\Big(2(2k_{\text{max}}+1)-1\Big)> 2(2k_{\text{max}}+1)-1=4k_{\text{max}}+1	
		\end{aligned},
		\label{ana_distance_B}
	\end{equation}\par	
	Taking the rapid decay of the sinc function into consideration, \eqref{ana_distance_B} inspires that the integral when $n\neq n^{\prime}$ is small enough to ignore.\par 
	Then we consider the situation when $n=n^{\prime}$. The integral interval is $[-t_{\text{max}},t_{\text{max}}]$. We have 
	\begin{equation}
		\scriptsize
		\begin{aligned}
			t_{\text{max}}-(x+\frac{B}{2})&=t_{\text{max}}-\frac{p_{i}\tau_{i}-n^{\prime}T}{1+p_{i}}-\frac{1}{2f_{\text{max}}(1+p_{i})}\\
			&> t_{\text{max}}-\tau_{i}-\frac{T}{2M(1+2k_{\text{max}})}> \frac{T}{M}\big(1-\frac{1}{2(1+2k_{\text{max}})}\big)>0\\
		\end{aligned},
		\label{ana_n=n'_right}
	\end{equation}
	and
	\begin{equation}
		\scriptsize
		\begin{aligned}
			(x-\frac{B}{2})-(-t_{\text{max}})&=\frac{p_{i}\tau_{i}-n^{\prime}T}{1+p_{i}}-\frac{1}{2f_{\text{max}}(1+p_{i})}+t_{\text{max}}\\
			&\geq t_{\text{max}}-\frac{n^{\prime}T}{1+p_{i}}-\frac{1}{2f_{\text{max}}(1+p_{i})}> \frac{T}{M}(l_{\text{max}}+1-1-\frac{1}{2(2k_{\text{max}}+1)})>0,
		\end{aligned}
		\label{ana_n=n'_left}
	\end{equation}
	which indicates that the entire mainlobe is contained. Moreover, similar to the analysis in \eqref{ana_distance_B}, we can obtain that the minimum distance $d$ between the margin of the mainlobe and the margin of the integral interval as $\frac{d}{B/2}\geq 4k_{\text{max}}+1$, which suggests that most of the active sidelobe is contained. As a result, we can replace the interval $[-t_{\text{max}},t_{\text{max}}]$ with $(-\infty,\infty)$ in \eqref{th2_eq_H_tf_DSE_singlepath} without noticeable error, which has a closed-form solution.\par 
	Now let us consider $t_{\text{max}}$ chosen in \eqref{system_constraints}. The inequality \eqref{ana_n=n'_right} is almost tight when $n^{\prime}=0$ and $\tau_{i}=\tau_{\text{max}}$, to make sure the correctness, $t_{\text{max}}>\tau_{\text{max}}$ must be guaranteed. Considering the delay-Doppler resolution, $t_{\text{max}}>\tau_{\text{max}}+\frac{T}{M}$ is a natural choice. At the meanwhile, if $t_{\text{max}}\geq\frac{T}{2}$, the inequality $(a)$ in \eqref{ana_n>n'} does not hold, which might cause $H^{i}_{n,m}[n^{\prime},m^{\prime}]$ for $n>n^{\prime}$ to be non-negligible and bring difficulty for the channel estimation and symbol detection design.\par 
	After the analysis of the relationship between $n$ and $n^{\prime}$, we obtain that $H^{i}_{n,m}[n^{\prime},m^{\prime}]\approx 0$ for $n\neq n^{\prime}$. Combining the Fourier transformation of sinc function with the analysis for $n=n^{\prime}$, we can derive that
	\begin{equation}
		\scriptsize
		H^{i}_{n,m}[n,m^{\prime}]\approx\left\{
		\arraycolsep=1pt\def\arraystretch{2.2}
		\begin{array}{lll}
			&\displaystyle \frac{\alpha_{i}}{1+p_{i}}e^{-j2\pi f\big(\frac{p_{i}\tau_{i}-nT}{1+p_{i}}\big)}, \ \ 
			&\displaystyle |f|\leq (1+p_{i})f_{\text{max}},\\
			&\displaystyle 0, \ \ 
			&\displaystyle \text{elsewhere}
		\end{array}
		\right.
		\label{H_n=n'_m_middle}
	\end{equation} 
	where we have $\alpha_{i}=\beta_{i}p_{i}e^{j2\pi p_{i}\tau_{i}\big(\nu_{i}-(m-m^{\prime})\Delta f\big)}$ and $f=f_c+m^{\prime}\Delta f-p_{i}(m-m^{\prime})\Delta f$ to simplify the notation. Now it is necessary to deliberate the relationship of $m$ and $m^{\prime}$.\par 
	If $m>m^{\prime}$, i.e., $m-m^{\prime}\geq 1$, we can infer that 
	\begin{equation}
		\scriptsize
		\begin{aligned}
			-(1+p_{i})f_{\text{max}}-f&=-(1+p_{i})f_{\text{max}}-f_c-m^{\prime}\Delta f+p_{i}(m-m^{\prime})\Delta f\\
			&\geq p_{i}\Delta f-f_c-m^{\prime}\Delta f-(1+p_{i})f_{\text{max}}\\
			&=p_{i}\Delta f(1-\frac{f_{c}}{p_{i}\Delta f}-\frac{m^{\prime}}{p_{i}}-\frac{(1+p_{i})f_{\text{max}}}{p_{i}\Delta f})\overset{(a)}{=}p_{i}\Delta f(1-\frac{\nu_{i}}{\Delta f}-\frac{m^{\prime}}{p_{i}}-\frac{(1+p_{i})f_{\text{max}}}{p_{i}\Delta f})\\
			&> p_{i}\Delta f(1-\frac{1}{4}-\frac{1}{N}-\frac{1}{2}-\frac{1}{2p_{i}})=p_{i}\Delta f(\frac{1}{4}-\frac{1}{N}-\frac{1}{2p_{i}})>0\\
		\end{aligned}.
		\label{ana_m>m'}
	\end{equation}\par
	If $m<m^{\prime}$, i.e., $m-m^{\prime}\leq-1$, we can derive that
	\begin{equation}
		\scriptsize
		\begin{aligned}
			f-(1+p_{i})f_{\text{max}}&=f_c+m^{\prime}\Delta f-p_{i}(m-m^{\prime})\Delta f-(1+p_{i})f_{\text{max}}\\
			&\geq f_c+m^{\prime}\Delta f+p_{i}\Delta f-(1+p_{i})f_{\text{max}}\geq (f_{c}-f_{\text{max}})+p_{i}(\Delta f-f_{\text{max}})>0
		\end{aligned}.
		\label{ana_m<m'}
	\end{equation}
	Considering \eqref{H_n=n'_m_middle}, it is feasible to ignore $H_{n,m}[n,m^{\prime}]$ when $m\neq m^{\prime}$. At the meantime, we have $f=f_{c}+m\Delta f>0$ for the case $m=m^{\prime}$. When it comes to the relationship of $f$ and the right margin, we can derive:
	\begin{equation}
		\scriptsize
		\begin{aligned}
			(1+p_{i})f_{\text{max}}-f&=f_{\text{max}}+\frac{f_{\text{max}}}{\nu_{i}}f_{c}-f_{c}-m\Delta f\overset{(a)}{>}f_{\text{max}}+2f_{c}-f_{c}-M\Delta f>f_{c}-M\Delta f>0\\
		\end{aligned},
		\label{ana_m=m'}
	\end{equation}
	which informs that the case $|f|\leq (1+p_{i})f_{\text{max}}$ in \eqref{H_n=n'_m_middle} can be applied. \par
	Now let us take a second look at the constraints in \eqref{system_constraints} especially for $f_{\text{max}}$. The representation in \eqref{ana_m>m'} $(a)$ tells that $f_{\text{max}}<\Delta f$ must be satisfied and the gap must be significant to save enough space for $\frac{\nu_{i}}{\Delta{f}}$. As a result, we have $f_{\text{max}}<\Delta{f}\ll M\Delta{f}$ for the inequality in \eqref{ana_m=m'} $(a)$. Meanwhile, taking the fact that $f_{c}>M\Delta f$ is always satisfied in realistic system, it is natural to choose $f_{\text{max}}>2\nu_{\text{max}}$. Similar to the choice of $t_{\text{max}}$, we require that $f_{\text{max}}>2\nu_{\text{max}}+\frac{\Delta f}{N}$. After that, \eqref{ana_m>m'} $(a)$ indicates that the constraint $f_{\text{max}}<\frac{\Delta f}{2}$ is nearly optimal.\par 
	In conclusion, we analyze the time-frequency coefficients and attain the approximation of $H_{n,m}[n^{\prime},m^{\prime}]$ when $\nu_{i}>0$ based on the system constraints presented in \eqref{system_constraints}.\par  The analysis when $p_{i}\leq 0$ is quite similar to the deduction before, where slight difference lies in \eqref{ana_n>n'} and \eqref{ana_n=n'_right}. $\frac{n^{\prime}T}{1+p_{i}}$ cannot be skipped directly since $1+p_{i}<0$ is inconsistent with the direction of inequality in \eqref{ana_n>n'}(a) and \eqref{ana_n=n'_right}(a), where $n<N$ is applied and the gap between $t_{\text{max}}$ and $\tau_{\text{max}}+\frac{T}{M}$ requires to be amplified. A convenient choice is $\tau_{\text{max}}+\frac{2T}{M}<t_{\text{max}}<\frac{T}{2}$, in which the following representation can be derived as:
	\begin{equation}
		\scriptsize
		\begin{aligned}
			H^{i}_{n,m}[n^{\prime},m^{\prime}]&\approx \beta_{i}\frac{|p_{i}|}{|1+p_{i}|}e^{j2\pi \tau_{i}f_{c}}e^{-j2\pi\frac{(m\Delta f+f_{c})(p_{i}\tau_{i}-nT)}{1+p_{i}}}\delta_{nn^{\prime}}\delta_{mm^{\prime}}
		\end{aligned}
		\label{H_tf_conclusion}.
	\end{equation}  
	Moreover, because no DSE is comprised if $\nu_{i}=0$, we have $H^{i}_{n,m}[n^{\prime},m^{\prime}]=\beta_{i}e^{-j2\pi \tau_{i}m\Delta f}\delta_{nn^{\prime}}\delta_{mm^{\prime}}$ when $\nu_{i}=0$, which can be treated as the limit for \eqref{H_tf_conclusion} when $p_{i}\rightarrow\infty$. To simplify the notation, we let $H^{i}[n,m]$ denote $H^{i}_{n,m}[n,m]$ and $H[n,m]=\sum_{i=1}^{N_{P}}H^{i}[n,m]$ in the following context. Since the approximation is precise enough through the former analysis, $\approx$ in \eqref{H_tf_conclusion} can also be replaced by $=$.\par
	\begin{figure}
		\center{\includegraphics[width=0.4\linewidth]{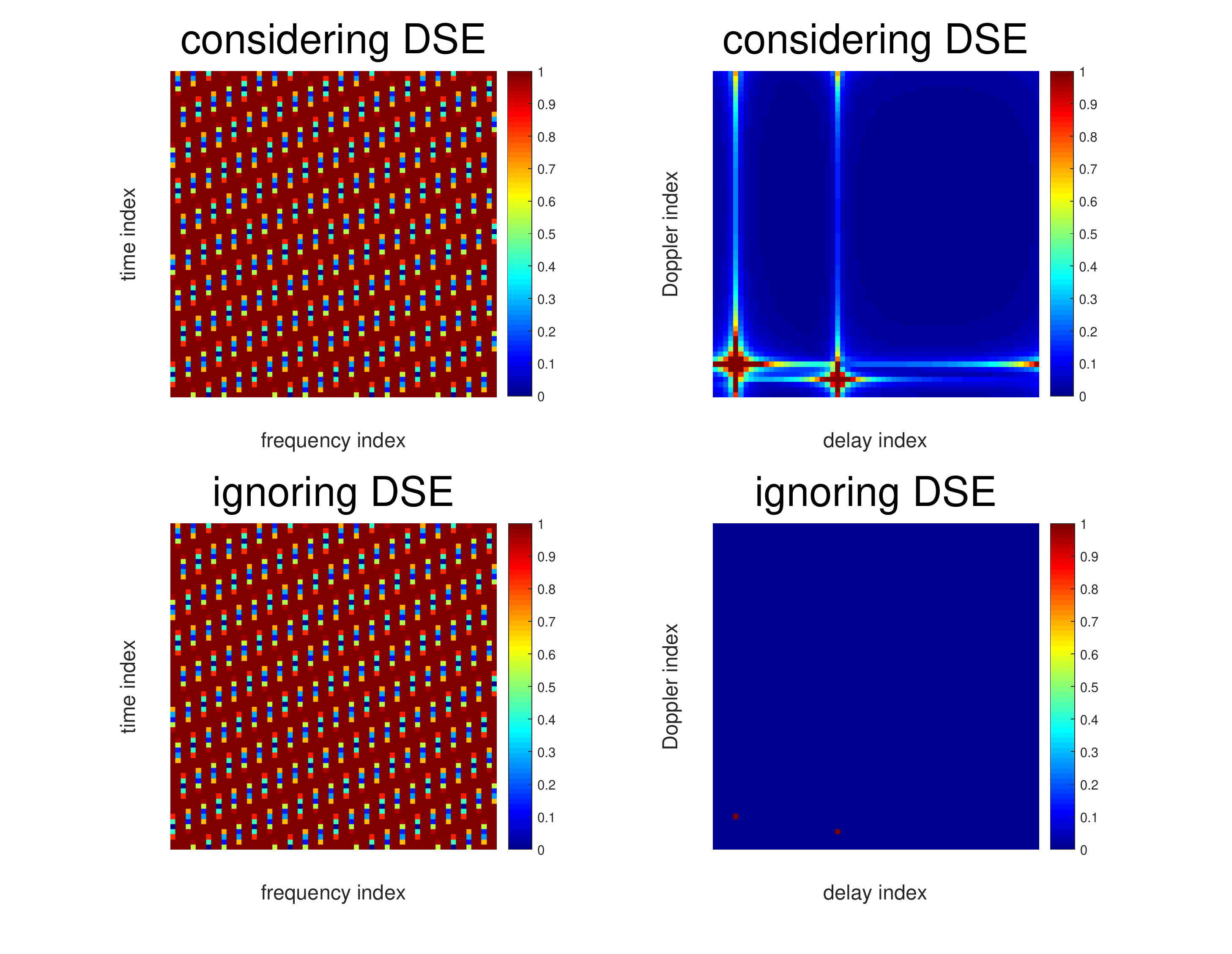}}
		\caption{Different channel coefficients in time-frequency domain and the delay-Doppler domain considering and ignoring DSE.}
		\label{Fig_channel_taomiu}	
	\end{figure}
	Fig. \ref{Fig_channel_taomiu} presents the difference brought by DSE by plotting the modulus of time-frequency channel coefficients $H[n,m]$ and delay-Doppler channel coefficients $h[k,l]$, where we set $N_{P}=2$ and only integer Doppler and delay are included. The modulus in time-frequency domain seems to have no visible change, however, significant difference appears in delay-Doppler domain. If ignoring DSE, there are only $N_{P}$ nonzeros in delay-Doppler channel coefficients corresponding to the Doppler index and delay index for each tap, which is imprecise since DSE occurs as long as $\nu_{i}\neq 0$. In DSE cases, though the most powerful channel coefficients lie in the same location, significant channel spread arises around the Doppler index and delay index for each tap. It is due to the time-frequency coupling brought by DSE, which destructs the widely-used sparse property and forces different schemes in channel estimation.\par
	\subsection{Factors Determining the Significance of DSE}
	In this subsection, the dominant factors that determines the significance of DSE will be addressed. \eqref{H_tf_conclusion} appears as the basic of this part, which is however too complicated to evaluate. In fact, we can make further analysis by applying some negligible approximations. Let $H_{o}^{i}[n,m]$ and $H^{i}_{s}[n,m]$ denote the time-frequency coefficients without and with considering DSE, respectively. $H_{o}^{i}[n,m]=\beta_{i}e^{j2\pi \tau_{i}\nu_{i}}e^{j2\pi\nu_{i}nT}e^{-j2\pi\tau_{i}m\Delta f}$ can be easily derived by substituting the $h(\tau,\nu)$ in \eqref{H_tf_general} with the sparse multipath channel model ignoring DSE directly. We can deduct that
	\begin{equation}
		\scriptsize
		\begin{aligned}
			\small
			H^{i}_{s}[n,m]&=\beta_{i}\frac{|p_{i}|}{|1+p_{i}|}e^{j2\pi \tau_{i}f_{c}}
			e^{-j2\pi\frac{(m\Delta f+f_{c})(p_{i}\tau_{i}-nT)}{1+p_{i}}}\\
			&\overset{(a)}{\approx}\beta_{i}e^{j2\pi \tau_{i}f_{c}}e^{-j2\pi f_{c}\tau_{i}\frac{p_{i}}{1+p_{i}}}e^{-j2\pi\frac{p_{i}m\Delta f\tau_{i}}{1+p_{i}}}e^{j2\pi\frac{m\Delta fnT}{1+p_{i}}}e^{j2\pi\frac{nTf_{c}}{1+p_{i}}}=\beta_{i}e^{j2\pi\tau_{i}\frac{f_{c}}{1+p_{i}}}e^{-j2\pi\frac{p_{i}m\Delta f\tau_{i}}{1+p_{i}}}e^{j2\pi\frac{m\Delta fnT}{1+p_{i}}}e^{j2\pi\frac{nTf_{c}}{1+p_{i}}}\\
			&\overset{(b)}{\approx}\beta_{i}e^{j2\pi\tau_{i}\nu_{i}}e^{-j2\pi \tau_{i}m\Delta f}e^{j2\pi\nu_{i}nT}e^{j2\pi \nu_{i}nT\frac{m\Delta f}{f_{c}}}\overset{(c)}{=}H_{o}^{i}[n,m]e^{j2\pi\nu_{i}\frac{m\Delta f}{f_{c}}nT}=H_{o}^{i}[n,m]e^{j2\pi mn\frac{\nu_{i}}{f_{c}}},
		\end{aligned}
		\label{H_tf_approx}
	\end{equation}
	where $(a)$ is obtained due to $|p_{i}|>10^{6}\gg 1$ and $(b)$
	is attained by applying $p_{i}=\frac{f_{c}}{\nu_{i}}$ and $|p_{i}|\gg 1$. $(c)$ is intuitive because it inspires us that DSE is embodied by the phase offset which is dependent on the Doppler offset and the time length.\par 
	From the representation in \eqref{H_tf_approx}, it is clear that the significance of DSE is independent of the ratio between the subcarrier spacing and the carrier frequency $\Delta f/f_{c}$ since the offset will be accumulated through the time-axis and the relationship $T\Delta f=1$ holds forever. It is different from the declaration in \cite{DSE_air}, where DSE is considered only due to the large bandwidth. Meanwhile, \eqref{H_tf_approx} provides the factors that determines the significance of DSE as follows:
	\begin{itemize}
		\item[1)] The size of OTFS symbol, which decides the maximum value of $m$ and $n$. DSE increases with $M$ and $N$ increasing;
		\item[2)] The mobility velocity, which affects the maximum value of $|\frac{\nu_{i}}{f_{c}}|$. DSE is significant when high-mobility is considered, which is the key problem OTFS aims to deal with.
	\end{itemize}\par 
	Taking the typical value $v=500$ km/h, $M=2048$ and $N=128$ as an example, DSE leads to a maximum phase offset about $e^{j0.24\pi}$, which is non-negligible especially in high-SNR scenarios. Unfortunately, DSE is independent of $\Delta{f}/f_{c}$, which has been declared in Section \ref{sec:intro} and reveals that DSE is required to consider in overall OTFS systems. To simplify the notation, we denote $\beta_{i}e^{j2\pi\tau_{i}\nu_{i}}$ as $\beta_{i}^{\prime}$ in the following part.\par
	\subsection{Delay-Doppler Domain Analysis}
	Since $Y[n,m]=H[n,m]X[n,m]$ can be satisfied from the analysis before, the delay-Doppler input-output can be approximately characterized by employing the properties of SFFT, which is provided by the following theorem.	
	\begin{theorem}
		\rm
		\label{th3_h_dd_bi}
		For bi-orthogonal pulses, input-output relationship in delay-Doppler domain can be represented as
		\begin{equation}
			\scriptsize
			y[k,l]=\sum_{i=1}^{N_{P}}\sum_{k^{\prime}=0}^{N-1}\sum_{l^{\prime}=0}^{M-1}h^{i}[(k-k^{\prime})_{N},(l-l^{\prime})_{M}]x[k^{\prime},l^{\prime}],
		\end{equation}
		where $h^{i}[k,l]$ can be formulated as \eqref{h_dd_bi_appro}.
		\begin{IEEEproof}
			The proof is provided in Appendix \ref{th3_proof_apendix}.
		\end{IEEEproof}
	\end{theorem}
	\begin{figure*}
		\begin{equation}
			\scriptsize
			h^{i}[k,l]\approx\beta_{i}^{\prime}e^{-j\pi(M-1)(\frac{l_{i}-l}{M})}e^{j\pi(N-1)(\frac{k_{i}-k}{N})}e^{j\pi\frac{(M-1)(N-1)}{2p_{i}}}\frac{\sin{\pi M(\frac{l_{i}-l}{M}-\frac{N-1}{2p_{i}})}}{M\sin{\pi(\frac{l_{i}-l}{M}-\frac{N-1}{2p_{i}})}}\frac{\sin{\pi N(\frac{k_{i}-k}{N}+\frac{M-1}{2p_{i}})}}{N\sin{\pi (\frac{k_{i}-k}{N}+\frac{M-1}{2p_{i}})}}
			\label{h_dd_bi_appro}
		\end{equation} 
		\hrulefill
	\end{figure*}
	$p_{i}$ in \eqref{h_dd_bi_appro} is ignored in traditional representation \cite{otfs_signal_model,OTFSmodel_cross} since $p_{i}$ is treated as $\infty$ compared with $MN$. However, this assumption will not hold true for OTFS systems according to the analysis before, which causes two modifications in \eqref{h_dd_bi_appro}:
	\begin{itemize}
		\item [1)] Delay-Doppler spread extension: Taking the integer delay for example, if the impact of $p_{i}$ is ignored, $h^{i}[k,l]=0,\forall l\neq l_{i}$ will hold true because of the zero point of sinc function. However, the extra phase introduced by $\frac{N-1}{2p_{i}}$ destructs this property and allows $h^{i}[k,l]\neq0,\forall l$, which can be shown in Fig. \ref{Fig_channel_taomiu}. Since $MN<|p_{i}|$ holds true for common scenarios, the most powerful channel coefficients still lie in the same location as the cases where DSE is ignored. However, significant channel spread arises around the Doppler index and delay index for each tap even though integer delay and Doppler can be achieved, which inspires the modification of channel estimation schemes ignoring DSE. 
		\item [2)] Extra phase shift modification: Besides the delay and Doppler spread extension, an extra phase shift $e^{j\pi\frac{(M-1)(N-1)}{2p_{i}}}$ is introduced for $\forall k,l$. It also deserves consideration to improve the reliability of phase-included alphabet such as the QPSK and 16QAM.   
	\end{itemize}\par 
	It is necessary to point out that approximation depicted in Appendix \ref{th3_proof_apendix} is employed to attain a closed-form representation of $h^{i}[k,l]$, whose precision can be verified by both the NMSE error and BER employing the approximated CSI in Fig. \ref{Fig_NMSE_diffM_perfect} and Fig. \ref{Fig_ber_perfect} in Section \ref{sec:simulation}.
	\section{OTFS With Rectangular Waveforms}
	\label{sec:recpulses}
	Since the ideal pulses satisfying the bi-orthogonal property cannot be realized in practice, we now provide the analysis of the OTFS systems with rectangular waveforms at both the transmitter and the receiver, where we have $g_{\text{rx}}(t)=g_{\text{tx}}(t)=\frac{1}{\sqrt{T}}\mathbb{I}_{[0,T]}(t)$.
	%\begin{equation}
	%	g_{\text{rx}}(t)=g_{\text{tx}}(t)
	%	\begin{cases}
	%		1/\sqrt{T},&t\in [0,T]\\
	%		0,&\text{elsewhere}\\
	%	\end{cases}.
	%\end{equation}\par
	The impact of noise is disregarded to simplify the notation like Section \ref{sec:system}. Similar to the analysis in \cite{otfs_signal_model,OTFS_crossdomain_reviewer4,OTFS_iterdetection_reviewer3,OTFS_MPCE_reviewer2} and so on, fractional delays are ignored by considering wideband systems design in this section, where $l_{i}$ is integer and $\tau_{i}\geq\frac{T}{M}$ can be derived for each path. Meanwhile, $l_{i}\leq M-1$ is assumed. As illustrated in Section \ref{sec:biopulses}, $|p_{i}|>10^{6}>MN$ is assumed to hold true consistently. A quick time-frequency domain analysis is first carried out to simplify the original representation in \eqref{H_tf_general}, which will be helpful for giving an approximated closed-form delay-Doppler input-output characterization. Meanwhile, since the scenario $\nu_{i}=0$ is trivial, we skip this analysis and assume that $|\nu_{i}|>0$ in this section.
	\subsection{Time-Frequency Domain Analysis}
	Since $g_{\text{rx}}(t)$ and $g_{\text{tx}}(t)$ are pulses on $[0,T]$, $A_{g_{\text{rx}},g_{\text{tx}}}(t,f)$ is nonzero for $-T<t<T$. Hence, the integral interval for $\tau$ in \eqref{H_tf_general} can be reduced to $[(n-n^{\prime}-1)T,(n-n^{\prime}+1)T]$. 
	Similar to the analysis in \cite{otfs_signal_model,OTFSmodel_cross}, the integral of the cross-ambiguity function can be approximated with a discrete sum as
	\begin{equation}
		\scriptsize
		A_{g_{\text{rx}},g_{\text{tx}}}(t,f)=\frac{1}{M}\sum_{q\in \mathcal{Q}}e^{-j2\pi f\frac{qT}{M}},
		\label{cross_appro_sum_rec}
	\end{equation}
	where we have 
	\begin{equation}
		\scriptsize
		\mathcal{Q}=
		\begin{cases}
			\{q\in\mathbb{N}:0\leq\frac{qT}{M}<t+T\},&t\in(-T,0)\\
			\{q\in\mathbb{N}:t\leq\frac{qT}{M}<T\},&t\in[0,T)\\
		\end{cases}.
		\label{cross_appro_sum_rec_range}
	\end{equation}\par
	The rectangular pulses do not satisfy the bi-orthogonal condition and generate ISI and ICI, which makes it difficult to carry out the channel estimation and equalization in the time-frequency domain. On the other hand, the data symbols are loaded in the delay-Doppler domain rather than the time-frequency domain. As a result, there is no need to study further in time-frequency domain. After the analysis before, we are already prepared to proceed further in a tractable manner in the delay-Doppler domain employing \eqref{cross_appro_sum_rec}, \eqref{cross_appro_sum_rec_range} and the approximation technique in bi-orthogonal waveforms.
	\subsection{Delay-Doppler Domain Analysis}
	By applying the SFFT of $Y[n,m]$ and substitute $h(\tau,\nu)$ and $A_{g_{\text{rx}},g_{\text{tx}}}(t,f)$ with the specified ones, the following theorem can be attained to describe the input-output relationship in delay-Doppler domain.
	\begin{theorem}
		\rm
		\label{th4_h_dd_rec}
		For rectangular pulses, input-output relationship in delay-Doppler domain can be represented as
		\begin{equation}
			\scriptsize
			y[k,l]=\sum_{i=1}^{N_{P}}\sum_{k^{\prime}=0}^{N-1}\sum_{l^{\prime}=0}^{M-1}h^{i}_{k,l}[k^{\prime},l^{\prime}]x[k^{\prime},l^{\prime}],
		\end{equation}
		where $h^{i}_{k,l}[k^{\prime},l^{\prime}]$ can be formulated as \eqref{h_dd_rec_eq}. The index sets are defined as
		\begin{equation}
			\scriptsize
			\mathcal{L}_{ISI}^{i}=
			\begin{cases}
				\{l^{\prime}\in \mathbb{N}:M-l_{i}+1\leq l^{\prime}\leq M-1\},&p_{i}>0\\
				\{l^{\prime}\in \mathbb{N}:M-l_{i}\leq l^{\prime}\leq M-1\},&p_{i}<0\\
			\end{cases}
		\end{equation}
		and
		\begin{equation}
			\scriptsize
			\mathcal{L}_{ICI}^{i}=
			\begin{cases}
				\{l^{\prime}\in \mathbb{N}:0\leq l^{\prime}\leq M-l_{i}\},&p_{i}>0\\
				\{l^{\prime}\in \mathbb{N}:0\leq l^{\prime}\leq M-l_{i}-1\},&p_{i}<0\\
			\end{cases},
		\end{equation}
		where $\mathcal{L}_{ICI}^{i}$ and $\mathcal{L}_{ISI}^{i}$ denote the $h^{i}_{k,l}[k^{\prime},l^{\prime}]$ from $H_{n,m}^{i}[n,m^{\prime}]$ and $H_{n,m}^{i}[n-1,m^{\prime}]$, respectively.
		\begin{IEEEproof}
			The proof is provided in Appendix \ref{th4_proof_apendix}.
		\end{IEEEproof}
	\end{theorem}
	\begin{figure*}
		\begin{equation}
			\scriptsize
			\label{h_dd_rec_eq}
			\begin{aligned}
				h^{i}_{k,l}[k^{\prime},l^{\prime}]&\approx\beta^{\prime}_{i}e^{-j\pi(M-1)(\frac{l_{i}+l^{\prime}-l}{M})}e^{j2\pi\nu_{i}\frac{l^{\prime}T}{M}}\\
				&\times
				\begin{cases}
					\frac{\sin{\pi M(\frac{l_{i}+l^{\prime}-l}{M}-\frac{N-2}{2p_{i}})}}{M\sin{\pi(\frac{l_{i}+l^{\prime}-l}{M}-\frac{N-2}{2p_{i}})}}\frac{\sin{\pi (N-1)(\frac{k_{i}+k^{\prime}-k}{N}+\frac{M-1}{2p_{i}})}}{N\sin{\pi (\frac{k_{i}+k^{\prime}-k}{N}+\frac{M-1}{2p_{i}})}}e^{j\pi (k_{i}+k^{\prime}-k)}e^{j\pi\frac{(N-2)(M-1)}{2p_{i}}}e^{-j2\pi\frac{k_{i}+k^{\prime}}{N}},& l^{\prime}\in\mathcal{L}_{ISI}^{i}\\
					\frac{\sin{\pi M(\frac{l_{i}+l^{\prime}-l}{M}-\frac{N-1}{2p_{i}})}}{M\sin{\pi(\frac{l_{i}+l^{\prime}-l}{M}-\frac{N-1}{2p_{i}})}}\frac{\sin{\pi N(\frac{k_{i}+k^{\prime}-k}{N}+\frac{M-1}{2p_{i}})}}{N\sin{\pi (\frac{k_{i}+k^{\prime}-k}{N}+\frac{M-1}{2p_{i}})}}e^{j\pi\frac{N-1}{N} (k_{i}+k^{\prime}-k)}e^{j\pi\frac{(N-1)(M-1)}{2p_{i}}},&l^{\prime}\in\mathcal{L}_{ICI}^{i}
				\end{cases}	
			\end{aligned}
		\end{equation}
		\hrulefill
	\end{figure*}
	Besides the delay-Doppler spread extension and extra phase shift in bi-orthogonal pulses, the modification due to DSE is embodied in the definition of $\mathcal{L}_{ISI}^{i}$ and $\mathcal{L}_{ISI}^{i}$ as well. Combining Theorem \ref{th4_h_dd_rec} with Theorem \ref{th3_h_dd_bi}, the delay-Doppler spread is actually the same for both the bi-orthogonal and rectangular pulses. The only difference is that the channel is shifted by an additional phase that depends on the delay-Doppler location, which is similar to the scenarios when DSE is ignored\cite{otfs_signal_model,OTFSmodel_cross}. However, the extended delay-Doppler spread still inspires us to reconsider the estimation schemes before where DSE is ignored.
	\section{OMP-based Channel Estimation Scheme}
	\label{sec:proposed}
	In this section, we will depict the OTFS channel estimation scheme based on the multipath channel presented in Theorem \ref{th1_dd_channel} which considers DSE. Since the prior estimation methods in \cite{OTFS_CE_threshold_classical,OTFS_MPCE_reviewer2,OTFS_CE_SBL} and so on consider only the sparse channel model where DSE is not included, it is necessary to develop new schemes to estimate the channel with DSE more precisely to improve the reliability of OTFS communication systems. Similar to the recent state-of-the-art analyses \cite{OTFS_CE_threshold_classical,OTFS_MPCE_reviewer2}, we first assumed OTFS systems are implemented with ideal waveforms for multipath channel considering DSE with integer delay and Doppler cases. Then the extension to the rectangular cases is naturally provided. The fractional delay can be ignored by considering typical wideband systems \cite{otfs_signal_model,wideband_integerdelay} while fractional Doppler shift can be estimated by employing off-grid algorithms such as Newtonized OMP (NOMP) \cite{NOMP_offgrid} and Sparse Bayesian Inference (SBI) \cite{SSBL_offgrid}, which inspires us to concentrate on the parameter estimation-based scheme considering only integer delay and Doppler shift.\par 
	\subsection{Problem Formulation}
	Impulse-based channel estimation technique is applied in this paper similar to the prior work \cite{OTFS_CE_threshold_classical,OTFS_CE_uplink1_OMP}, where the pilot is allocated as
	\begin{equation}
		\scriptsize
		\label{pilot_frame}
		x[k,l]=
		\begin{cases}
			x_{p},&k=0,l=0\\
			0,&\text{elsewhere}
		\end{cases}.
	\end{equation}
	The received signal can be derived as 
	\begin{equation}
		\scriptsize
		Y[n,m]=\frac{x_{p}}{\sqrt{NM}}H[n,m]+V[n,m],
		\label{receive_tf_cs}
	\end{equation}
	where $V[n,m]\sim\mathcal{CN}(0,\sigma^{2})$ denotes the additive white
	Gaussian noise (AWGN) samples in the Time-Frequency domain. From \eqref{H_tf_approx}, the time-frequency channel coefficients can be rewritten as follows:
	\begin{equation}
		\scriptsize
		H[n,m]=\sum_{k=-k_{\text{max}}}^{k_{\text{max}}}\sum_{l=0}^{l_{\text{max}}}\beta[k,l]e^{-j2\pi \frac{ml}{M}}e^{j2\pi \frac{kn}{N}}e^{j2\pi \frac{mnk\Delta f}{Nf_{c}}}	,
		\label{H_tf_sparse_repre}
	\end{equation}
	where $\beta[k,l]$ is equal to $\beta_{k,l}e^{j2\pi\frac{kl}{NM}}$ and $\beta[k,l]=0$ means that there is no such path as $\nu_{i}=\frac{k}{N}\Delta f$ and $\tau_{i}=\frac{l}{M}T$. Let $\mathbf{y}_{\text{tf}}\in\mathbb{C}^{NM\times 1}$, $\boldsymbol{\beta}\in \mathbb{C}^{ (2k_{\text{max}}+1)(l_{\text{max}}+1)\times 1}$ and $\mathbf{v}_{\text{tf}}\sim\mathcal{CN}(\mathbf{0},\sigma^{2}\mathbf{I}_{MN})$ denote the vectorized $Y[n,m]$, $\beta[k,l]$ and time-frequency noise $V[n,m]$, we can derive 
	\begin{equation}
		\scriptsize
		\mathbf{y}_{\text{tf}}=\frac{x_{p}}{\sqrt{NM}}\mathbf{\Phi}_{\text{tf}} \boldsymbol{\beta} +\mathbf{v}_{\text{tf}}
		\label{receive_tf_vector},
	\end{equation}
	where $\mathbf{\Phi}_{\text{tf}}\in \mathbb{C}^{NM\times (2k_{\text{max}}+1)(l_{\text{max}}+1)}$ is the sensing matrix whose elements can be formulated as follows:
	\begin{equation}
		\scriptsize
		\mathbf{\Phi}_{\text{tf}}(n+mN,k+ll_{\text{max}})=e^{-j2\pi \frac{ml}{M}}e^{j2\pi \frac{kn}{N}}e^{j2\pi \frac{mnk\Delta f}{Nf_{c}}}.
	\end{equation}
	$\mathbf{\Phi}_{\text{tf}}(:,k+ll_{\text{max}})$ is regarded as the channel basis with respect to Doppler tap $\frac{k}{N}\Delta f$ and delay tap $\frac{l}{M}T$.
	\par 
	Similarly, we can rewrite the received signal $\mathbf{y}_{\text{dd}}$ in delay-Doppler domain as
	\begin{equation}
		\scriptsize
		\mathbf{y}_{\text{dd}}=x_{p}\mathbf{\Phi}_{\text{dd}} \boldsymbol{\beta} +\mathbf{v}_{\text{dd}},
	\end{equation}
	where $\mathbf{\Phi}_{\text{dd}}$ can be obtained by Theorem \ref{th3_h_dd_bi} or  \eqref{th3_proof_eq}(a) for more accurate results. $\mathbf{v}_{\text{dd}}\sim\mathcal{CN}(\mathbf{0},\sigma^{2}\mathbf{I}_{MN})$ is the additive white Gaussian noise due to the property of SFFT. Let $c_{\text{dd}}=x_{p}$ and $c_{\text{tf}}=x_{p}/\sqrt{MN}$, the channel estimation problem can therefore be formulated as a sparse recovery problem as follows:
	\begin{equation}
		\scriptsize
		\begin{aligned}
			\min_{\boldsymbol{\beta}}\quad&||\boldsymbol{\beta}||_{0}\quad\\
			\quad \text{s.t.}\quad &||\mathbf{y}_{\alpha}-c_{\alpha}\mathbf{\Phi}_{\alpha}\boldsymbol{\beta}||_{2}<\epsilon,
		\end{aligned}
		\label{problem_cs}
	\end{equation}
	where $\alpha\in\{\text{dd},\text{tf}\}$ denotes the choice that the channel estimation is carried out in which domain. Besides, the representation in \eqref{H_tf_conclusion} can be also applied, in which the only difference is the sensing matrix $\mathbf{\Phi}_{\alpha}$. In fact, replacing \eqref{H_tf_conclusion} with \eqref{H_tf_approx} caused almost no error because $|p_{i}|\gg 1$, which will be confirmed again in simulation results.\par 
	\subsection{OMP-Based Channel Estimation Scheme}
	Since the channel estimation problem has been concluded to a standard sparse signal recovery problem in \eqref{problem_cs}, various low-complexity greedy algorithms such as orthogonal matching pursuit (OMP) \cite{OMP_ref} and subspace pursuit (SP) \cite{SP_ref} can be directly employed to recover the channel. In this subsection, we provide the channel estimation scheme based on the classic OMP algorithm.
	\begin{algorithm}  
		\renewcommand{\algorithmicrequire}{\textbf{Input:}}
		\renewcommand{\algorithmicensure}{\textbf{Output:}}
		\caption{OMP-based OTFS channel estimation considering DSE}
		\label{alg:1}
		\scriptsize
		\begin{algorithmic}[1]
			\REQUIRE
			$\mathbf{y}=\frac{1}{c_{\alpha}}\mathbf{y}_{\alpha}$, the sensing matrix $\mathbf{\Phi}_{\alpha}$
			\ENSURE
			the estimated channel vector $\hat{\mathbf{h}}_{\alpha}$
			\STATE
			initialize $\boldsymbol{\beta}=\mathbf{0}$, $\mathcal{S}=\varnothing$, $\mathbf{r}=\mathbf{y}$
			\REPEAT
			\STATE $\mathbf{\Psi}=\mathbf{\Phi}_{\alpha}^{H}\mathbf{r}$,
			\STATE
			$\hat{q}=\mathop{\arg\max}_{q}|\mathbf{\Psi}(q)|$
			\STATE 
			$\mathcal{S}=\mathcal{S}\cup\{\hat{q}\}$
			\STATE 
			$\boldsymbol{\beta}_{\mathcal{S}}=\mathbf{\Phi}_{\alpha,\mathcal{S}}^{\dagger}\mathbf{y}$
			\STATE
			$\mathbf{r}=\mathbf{y}-\mathbf{\Phi}_{\alpha,\mathcal{S}}\boldsymbol{\beta}_{\mathcal{S}}$
			\UNTIL
			stopping criteria
			\STATE	
			return $\hat{\mathbf{h}}_{\alpha}=\mathbf{\Phi}_{\alpha}\boldsymbol{\beta}$
		\end{algorithmic}		
	\end{algorithm}   \par 
	As illustrated in \textbf{Algorithm \ref{alg:1}}, we treat the normalized received signal $\mathbf{y}$, the sensing matrix $\mathbf{\Phi}_{\alpha}$ as the input. In the $i^{th}$ iteration, we compute the inner dot between the residue of the received signal and each column of the sensing matrix $\mathbf{\Phi}_{\alpha}$, in which we determine the path including the highest correlation and add this path to the current path support $\mathcal{S}$. After that, the nonzero values in $\boldsymbol{\beta}$ are updated by the least square method to achieve OMP. At last, the residue is computed again so that the previous effects are removed. The iteration is terminated when stopping criteria established in advance is satisfied, e.g., $||\mathbf{r}||_2<\epsilon$ or the maximum iteration times are reached. The estimation results of $\boldsymbol{\beta}$ and the corresponding channel $\hat{\mathbf{h}}_{\alpha}=\mathbf{\Phi}_{\alpha}\boldsymbol{\beta}$ are attained then, which can be employed to detect the transmitted data. \par 
	The major computational complexity lies in $\boldsymbol{\beta}_{\mathcal{S}}=\mathbf{\Phi}_{\alpha,\mathcal{S}}^{\dagger}\mathbf{y}$, whose complexity can be bounded by $\mathcal{O}(N_{P}^{2}MN)$. On the other hand, the maximum iteration times cater to the sparsity $N_{P}$ in general\cite{OMP_ref}. So the total complexity can be bounded by $\mathcal{O}(N_{P}^{3}MN)$, which is feasible since $N_{P}$ is usually a small number such as $4$ and the SFFT itself takes up complexity more than $MN\log{MN}$ to recover the data.\par
	It is also meaningful to describe the possible promotion of the channel estimation scheme depicted before. In fact, the complexity might be diminished by employing the ``sparsity" in delay-Doppler domain presented in Fig. \ref{Fig_channel_taomiu}, which inspires us that the sensing matrix $\mathbf{\Phi}_{\text{dd}}$ includes few nonzeros, though the sparsity here is different from the sparsity widely employed in prior works. Moreover, the problem of the fractional Doppler can be almost fully settled by improving the resolution of $\mathbf{\Phi}_{\alpha}$. At last, more elaborate design of the pilot considering the block sparsity under DSE might be helpful for joint pilot-data transmission, which is a promising prospect to promote the feasibility of OTFS systems.
	\subsection{Extension to OTFS with Rectangular waveforms} 
	So far, we have focused on the channel estimation scheme for OTFS systems with bi-orthogonal waveforms. Though the ideal pulses cannot be realized in practical communication systems, the channel estimation scheme illustrated in \textbf{Algorithm \ref{alg:1}} is still available, where we only require to replace the delay-Doppler sensing matrix $\mathbf{\Phi}_{\text{dd}}$ according to Theorem \ref{th4_h_dd_rec} and the pilot frame structure in \eqref{pilot_frame}.
	\section{Simulation Results}
	\label{sec:simulation}
	In this section, the significance of DSE and the performance of proposed channel estimation scheme will be evaluated by simulation results. The typical value of relevant simulation parameters is provided in Table \ref{simulation_para_table}. The complex gain of each path is randomly generated as $\beta_{i}\sim\mathcal{CN}(0,1/N_{P})$. Moreover, since SFFT and ISFFT are both unitary transformations, channel estimation and linear equalization in delay-Doppler domain is equivalent to those in the time-frequency domain. As a result, the time-frequency low-complexity LMMSE-based equalization technique \cite{low_complex_data_detection} is applied in this section as
	\begin{equation}
		\scriptsize
		\hat{X}[n,m]=\frac{H^{*}(n,m)Y[n,m]} {|H[n,m]|^{2}+\sigma^{2}/ \sigma_{s}^{2} }
		\label{tf_equalization},
	\end{equation}
	where $\sigma_{s}^{2}$ and $\sigma^{2}$ denote the average power of time-frequency symbols and noise, respectively.\par
	\begin{table}
		\caption{Simulation Parameters}
		\centering
		\label{simulation_para_table}
		\renewcommand\arraystretch{1.2}
		\begin{tabular}{p{15em}p{10em}}
			\hline
			Parameter &
			Typical value\\
			\hline
			Carrier frequency ($f_{c}$)& $4$GHz\\
			Subcarrier spacing ($\Delta f$)& $15$kHz\\
			Number of subcarriers ($M$)& 512 (128$\sim$2048)\\
			Number of time slots ($N$)& 128\\
			UE speed (km/h)& 100,360,500\\
			Modulation alphabet & QPSK,16QAM\\
			Maximum delay grid ($l_{\text{max}}$)& 20\\
			Number of paths ($N_{P}$)& 4\\
			\hline
		\end{tabular}
	\end{table}
	\subsection{Significance of DSE}
	We first demonstrate the essentiality of considering DSE by presenting the deviation in the normalized mean square error (NMSE) of the delay-Doppler channel $\mathbf{h}_{\text{dd}}$ and the bit-error-rate (BER) performance. The approximation precision of Theorem \ref{th3_h_dd_bi} is also evaluated in this subsection. $\mathbf{h}_{\text{dd}}$ is obtained by carrying out the SFFT of \eqref{H_tf_conclusion}. The NMSE is defined as 
	\begin{equation}
		\scriptsize
		\text{NMSE}=\mathbb{E}\frac{||\mathbf{h}_{\text{dd}}-\hat{\mathbf{h}}_{\text{dd}}||_{2}^{2}}{||\mathbf{h}_{\text{dd}}||_{2}^{2}},
		\label{NMSE_repre}
	\end{equation}
	where $\hat{\mathbf{h}}_{\text{dd}}$ can denote the vectorized channel coefficients ignoring DSE or considering approximated DSE. Perfect knowledge of the channel parameters including the time-delay, Doppler shift and complex gain of each path is assumed in this subsection.\par 
	\begin{figure}
		\center{\includegraphics[width=0.4\linewidth]{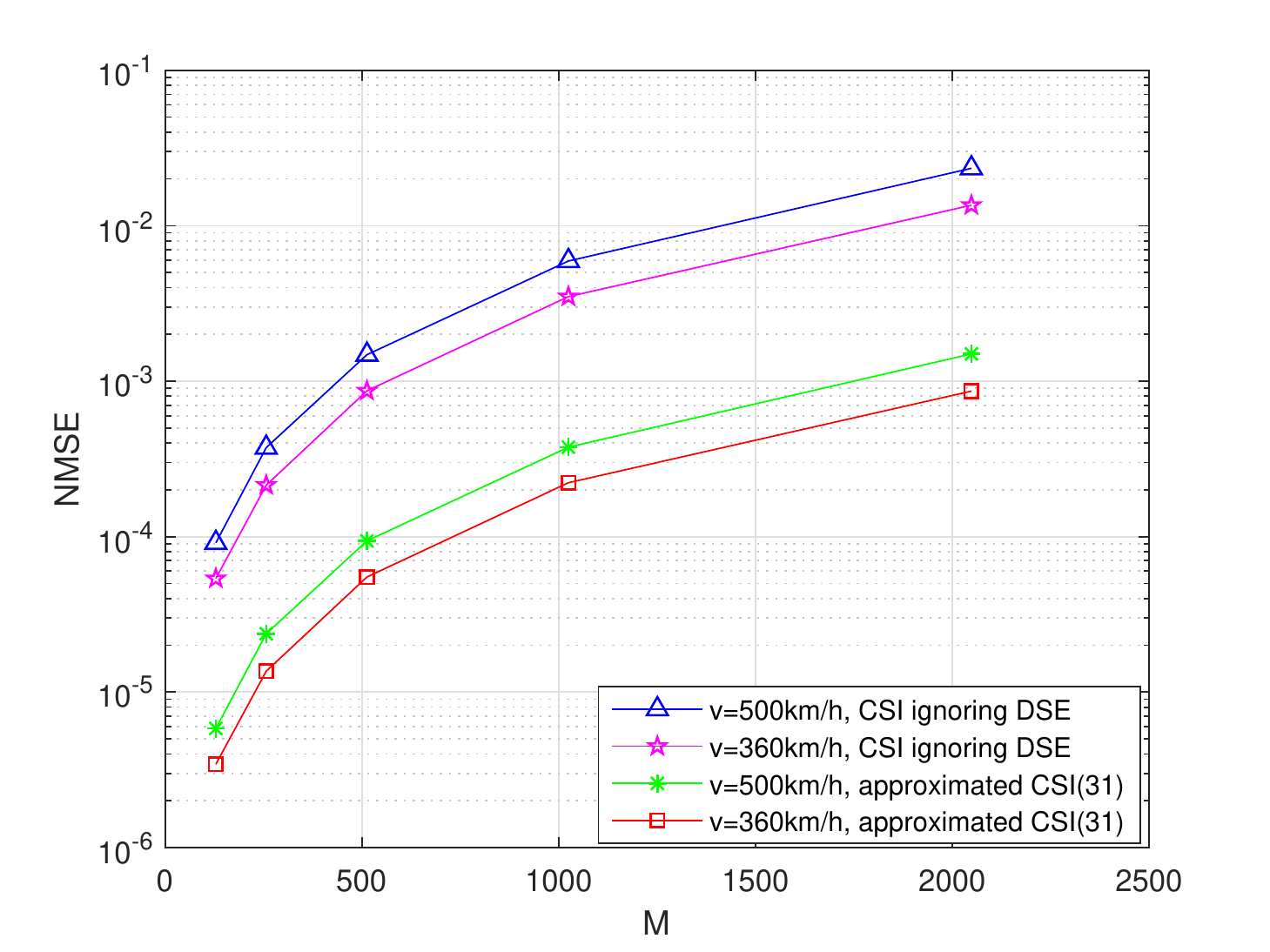}}
		\caption{NMSE against $M$ with perfect knowledge of the channel parameters.}
		\label{Fig_NMSE_diffM_perfect}	
	\end{figure}
	In Fig. \ref{Fig_NMSE_diffM_perfect}, the NMSE of delay-Doppler channel vector against $M$ under different receiver velocities is explored. It is clearly that DSE increases with $M$ and $v$ increasing, which confirms the analysis before. When it comes to the situation at $M=2048$ and $v=500$ km/h, the NMSE becomes more than 2\%, which brings substantial decoding error if DSE is ignored especially in high SNR scenarios. The approximation precision for our closed-form representation is also presented in Fig. \ref{Fig_NMSE_diffM_perfect}, where we find the NMSE between the approximated CSI in \eqref{h_dd_bi_appro} and the accurate CSI is less than 10\% of the NMSE between the CSI ignoring DSE and the accurate one. \par 
	\begin{figure}
		\center{\includegraphics[width=0.4\linewidth]{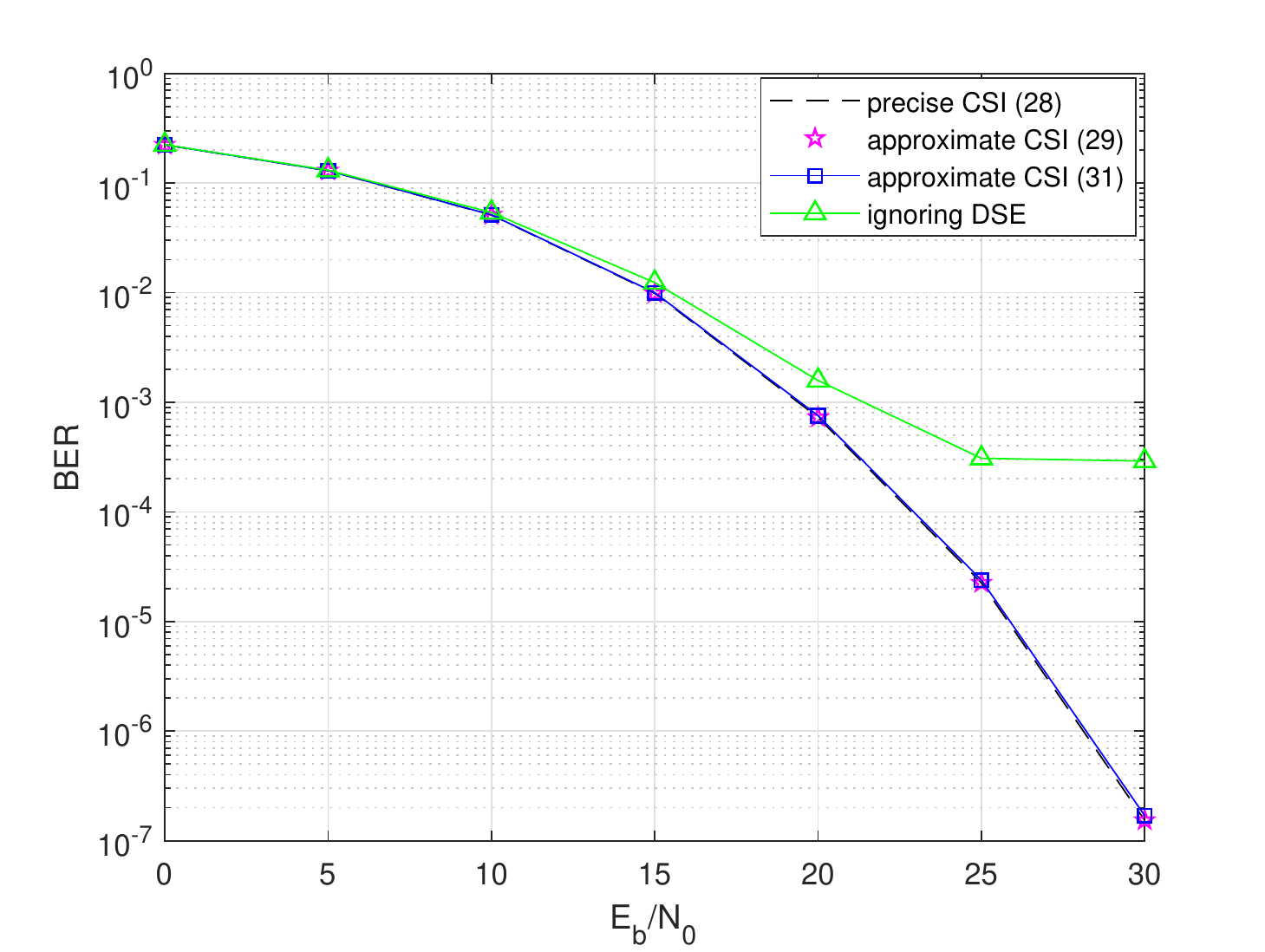}}
		\caption{BER performance against $E_{b}/N_{0}$ with perfect knowledge of the channel parameters and the standard 16QAM alphabet.}
		\label{Fig_ber_perfect}	
	\end{figure}
	Fig. \ref{Fig_ber_perfect} demonstrates the importance of considering DSE by plotting the BER performance against the data SNR $E_{b}/N_{0}$. 16QAM is selected as the modulation alphabet and the number of subcarriers is set as $M=512$. It is clear that when $E_{b}/N_{0}>25$dB, the error floor occurs due to the inaccuracy of channel coefficients ignoring DSE. However, we can diminish the error floor whose level is approximately $3\times 10^{-4}$ by considering DSE, which is helpful for the utilization in realistic systems. Moreover, from the BER performance, no difference is revealed if we replace the near-accurate representation in \eqref{H_tf_conclusion} with the approximation in \eqref{H_tf_approx} and even \eqref{h_dd_bi_appro}, which provides enough convenience to simplify the computation. Based on this observation, we employ \eqref{H_tf_approx} to generate the sensing matrix rather than \eqref{H_tf_conclusion} in the following simulation.\par
	\subsection{Performance of Proposed Channel Estimation Scheme}
	After illustrating the importance of considering DSE, the performance of our proposed channel estimation scheme is assessed by both NMSE and BER performance. The definition of NMSE is similar to \eqref{NMSE_repre}, where $\hat{\mathbf{h}}_{\text{dd}}$ denotes the estimated delay-Doppler CSI. The widely-used threshold-based technique \cite{OTFS_CE_threshold_classical} which can be seen as the OMP-based estimation scheme ignoring DSE is selected as the comparison, where the threshold is set as $3\sigma$. The circulation in \textbf{Algorithm \ref{alg:1}} is terminated when iteration times reach $N_{P}$. It is worth pointing out that in ultra-high SNR circumstances, both methods can achieve no estimation loss of the channel coefficients, however, the physic explanation of the estimated results is different, in which the threshold-based scheme obtains a channel without sparse characteristic. The pilot SNR is defined as $\text{SNR}_{p}=|x_{p}|^{2}/\sigma^{2}$ consistent with \cite{OTFS_MPCE_reviewer2,OTFS_CE_threshold_classical,OTFS_CE_Windowdesign,OTFS_CE_uplink1_OMP}. Though $\text{SNR}_{p}$ is usually a high value such as 45 dB, ISFFT will spread the power into $MN$ time-frequency grids uniformly, which leads to an extremely low $\text{SNR}=\frac{|x_{p}|^{2}}{MN\sigma^{2}}$ when it comes to the average power on each time-frequency grid, e.g., $\text{SNR}_{p}=45$dB leads to the average $\text{SNR}$ which are less than $0$dB when $M=512$ and $N=128$ in time-frequency domain.\par  
	\begin{figure}
		\center{\includegraphics[width=0.4\linewidth]{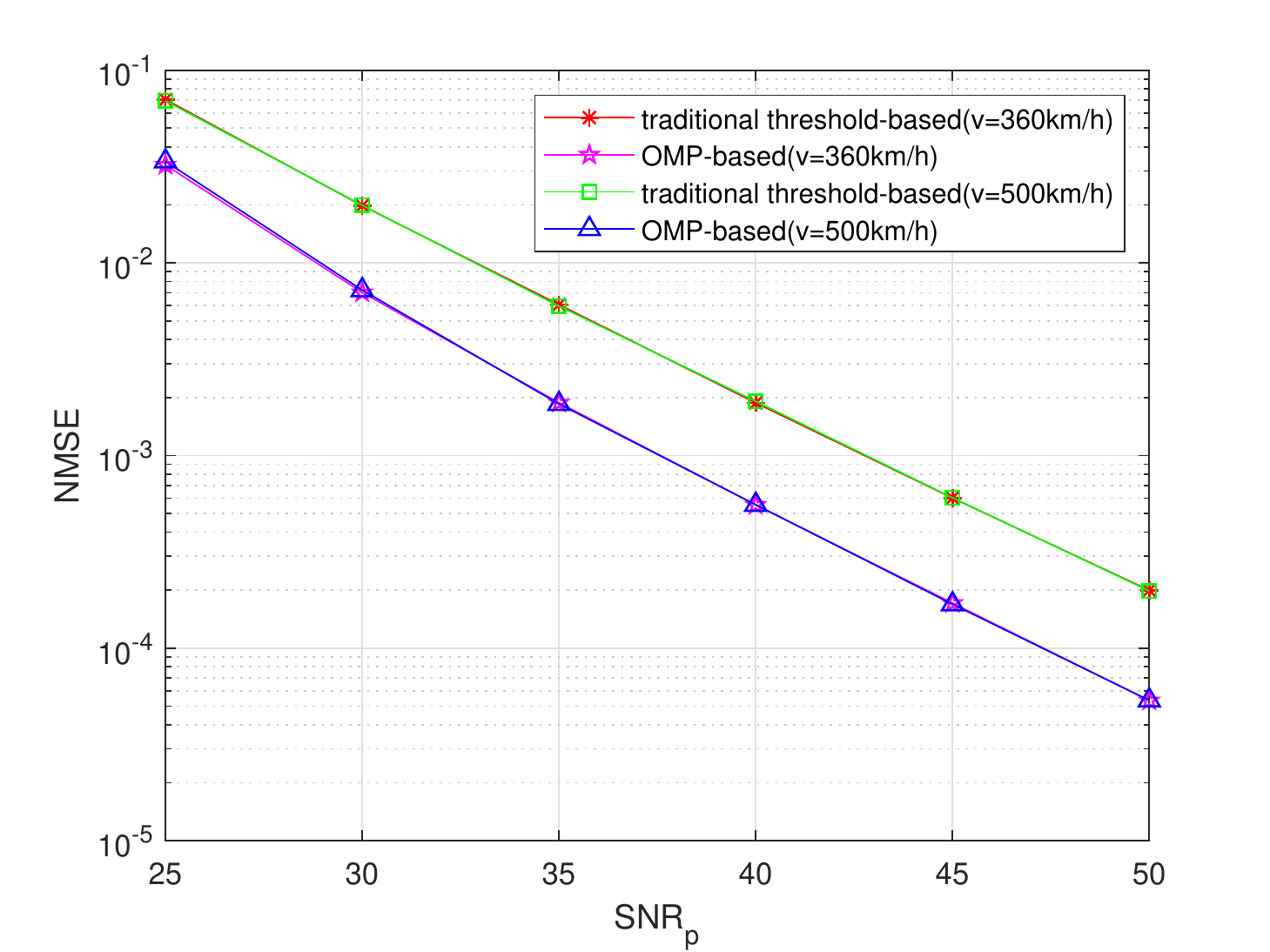}}
		\caption{NMSE performance of the channel estimation against $\text{SNR}_{p}$ with $N=64$ and $M=128$.}
		\label{Fig_NMSE_CE_64_128}	
	\end{figure}
	Fig. \ref{Fig_NMSE_CE_64_128} presents the NMSE performance comparison against $\text{SNR}_{p}$ under different velocities when $M=128$ and $N=64$. The user velocity has little impact on the channel estimation performance because OTFS is a technique including 2D-modulation. Our proposed scheme can achieve a NMSE less than $6\times 10^{-4}$ when $\text{SNR}_{p}>40$dB. Moreover, in high SNR scenarios, the proposed OMP-based technique outperforms the traditional threshold-based scheme by more than $5$dB.\par 
	\begin{figure}
		\center{\includegraphics[width=0.4\linewidth]{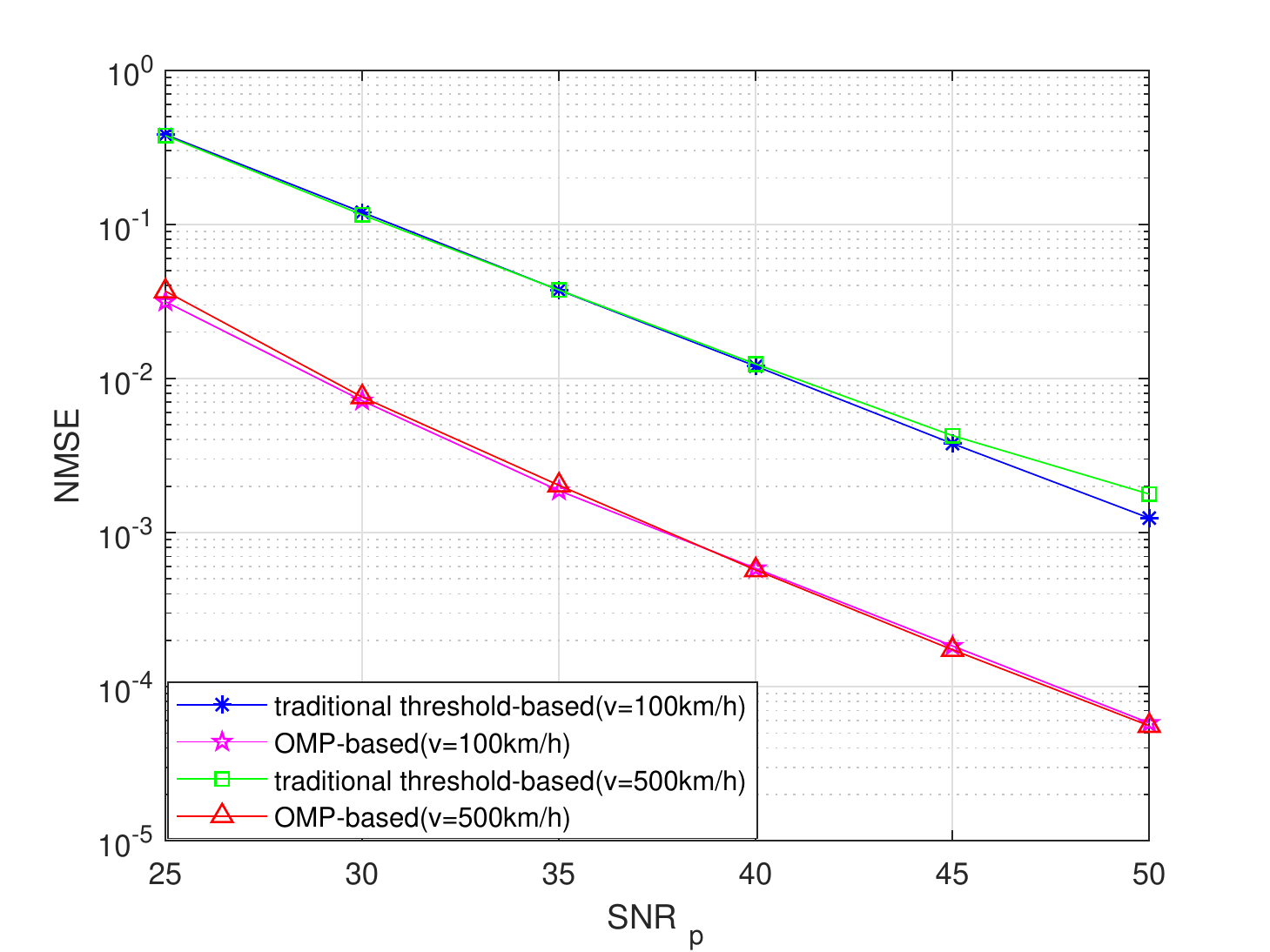}}
		\caption{NMSE performance of the channel estimation against $\text{SNR}_{p}$ with $N=128$ and $M=512$.}
		\label{Fig_NMSE_CE_128_512}	
	\end{figure}
	In Fig. \ref{Fig_NMSE_CE_128_512}, we show NMSE performance comparison against $\text{SNR}_{p}$ under different velocities when $M=512$ and $N=128$. The SNR precedence of proposed OMP-based method is amplified to about $12$dB, which is appreciable enough to deserve the consideration. The superiority is owing to larger DSE brought by larger $M$ and $N$, which offers more benefits by developing channel estimation schemes taking DSE into consideration. At the meantime, OMP-based channel estimation scheme can work without knowledge of the noise variance, which is a key parameter in traditional threshold-based method.\par
	\begin{figure}
		\center{\includegraphics[width=0.4\linewidth]{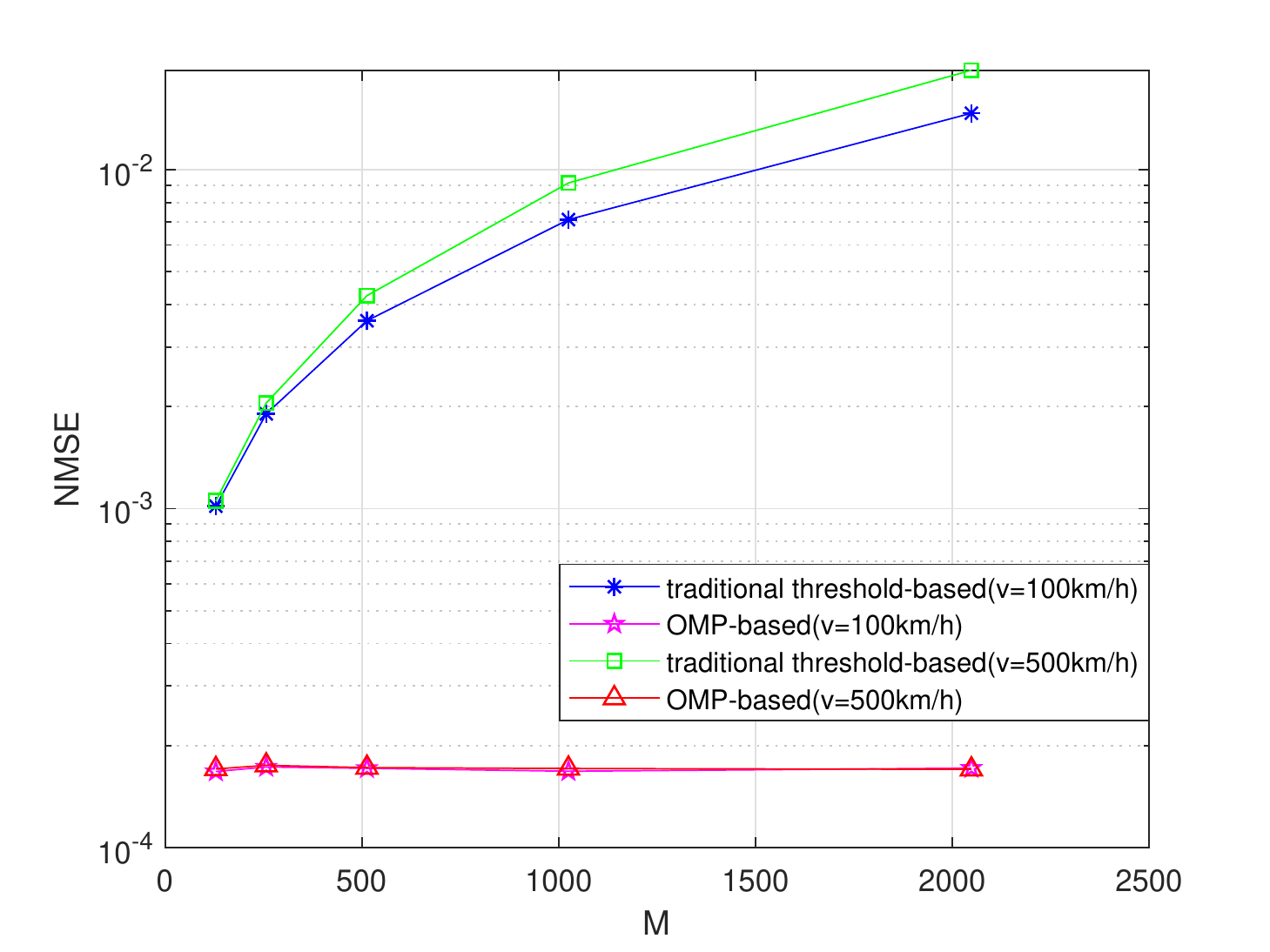}}
		\caption{NMSE performance of the channel estimation against $M$ with $N=128$ and $\text{SNR}_{p}=45$dB.}
		\label{Fig_NMSE_45dB_128}	
	\end{figure}
	Fig. \ref{Fig_NMSE_45dB_128} displays NMSE comparison against $M$ under different velocities with $N=128$ and $\text{SNR}_{p}=45$dB. NMSE of proposed OMP-based scheme is less than $2\times 10^{-4}$ and stays uncorrelated from $M$. However, due to the ever-increasing DSE, NMSE of threshold-based method increases as $M$ and $v$ boosts, in which NMSE approaches $1\times 10^{-2}$ when $M=1024$. It is unbearable for OTFS systems since $M$ and $N$ is required to be large enough to defend the doubly-dispersive channel.\par 
	\begin{figure}
		\center{\includegraphics[width=0.4\linewidth]{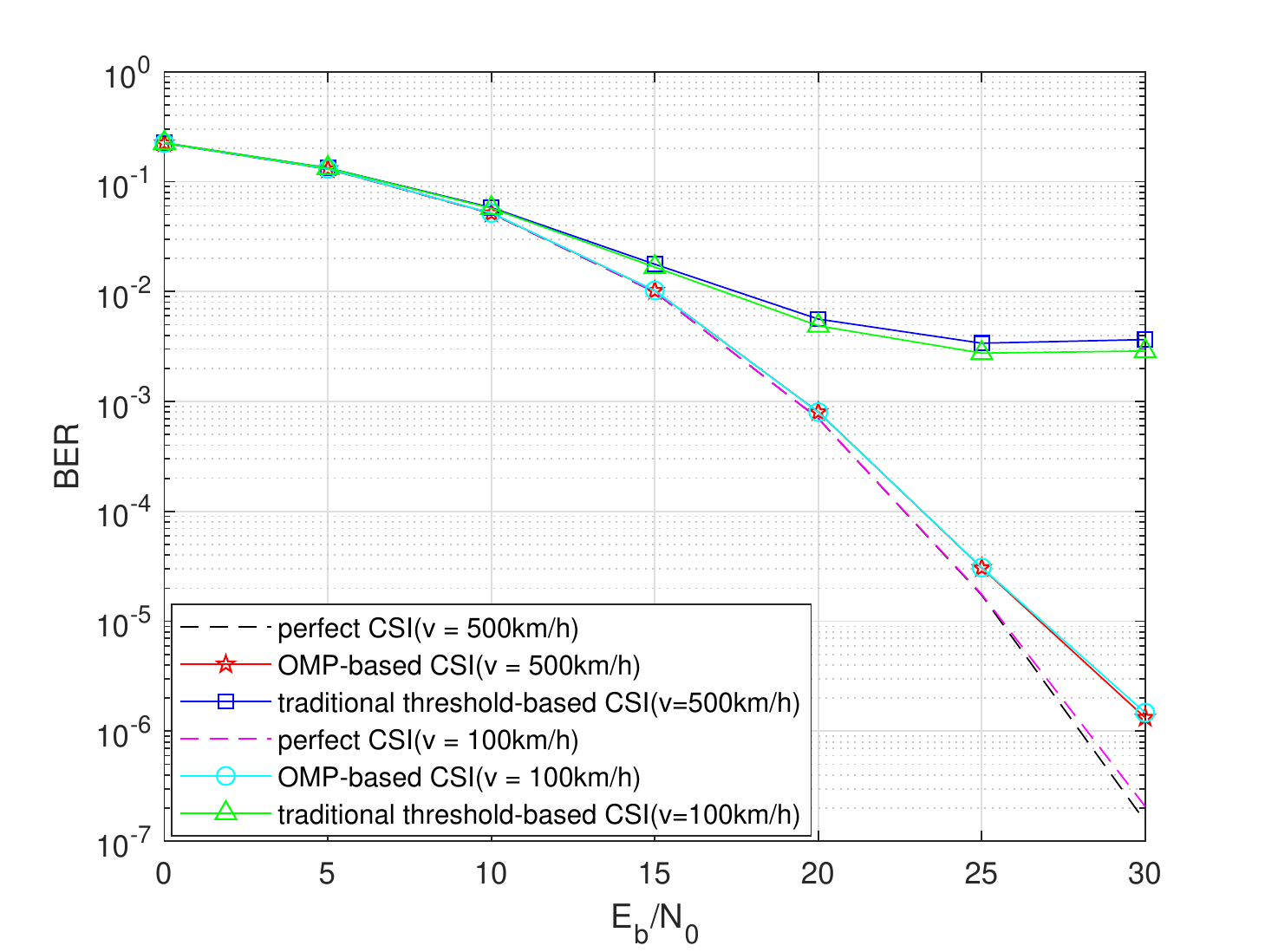}}
		\caption{BER performance under different UE velocity caused by different channel estimation scheme against $E_{b}/N_{0}$ with $N=128$, $M=512$, $\text{SNR}_{p}=45$dB and the standard 16QAM alphabet.}
		\label{Fig_BER_45dB_diffvelocity}	
	\end{figure}
	Since the target of channel estimation is to improve the reliability of data detection, it also makes sense to present BER performance employing the estimated CSI based on different channel estimation schemes. In Fig. \ref{Fig_BER_45dB_diffvelocity}, we show BER comparison against data SNR $E_{b}/N_{0}$ under different velocities. $\text{SNR}_{p}$ is set as $45$dB and the modulation alphabet we choose is 16QAM. BER floor more than $2\times 10^{-3}$ occurs due to the inaccurate CSI based on the threshold-based method, which reveals the deficiency of channel estimation approaches ignoring DSE. However, if OMP-based CSI considering DSE is acquired, BER can be smaller than $1\times 10^{-4}$ when $E_{b}/N_{0}>25$dB, which demonstrates the excellent performance of our proposed estimation scheme again. \par  
	\begin{figure}
		\center{\includegraphics[width=0.4\linewidth]{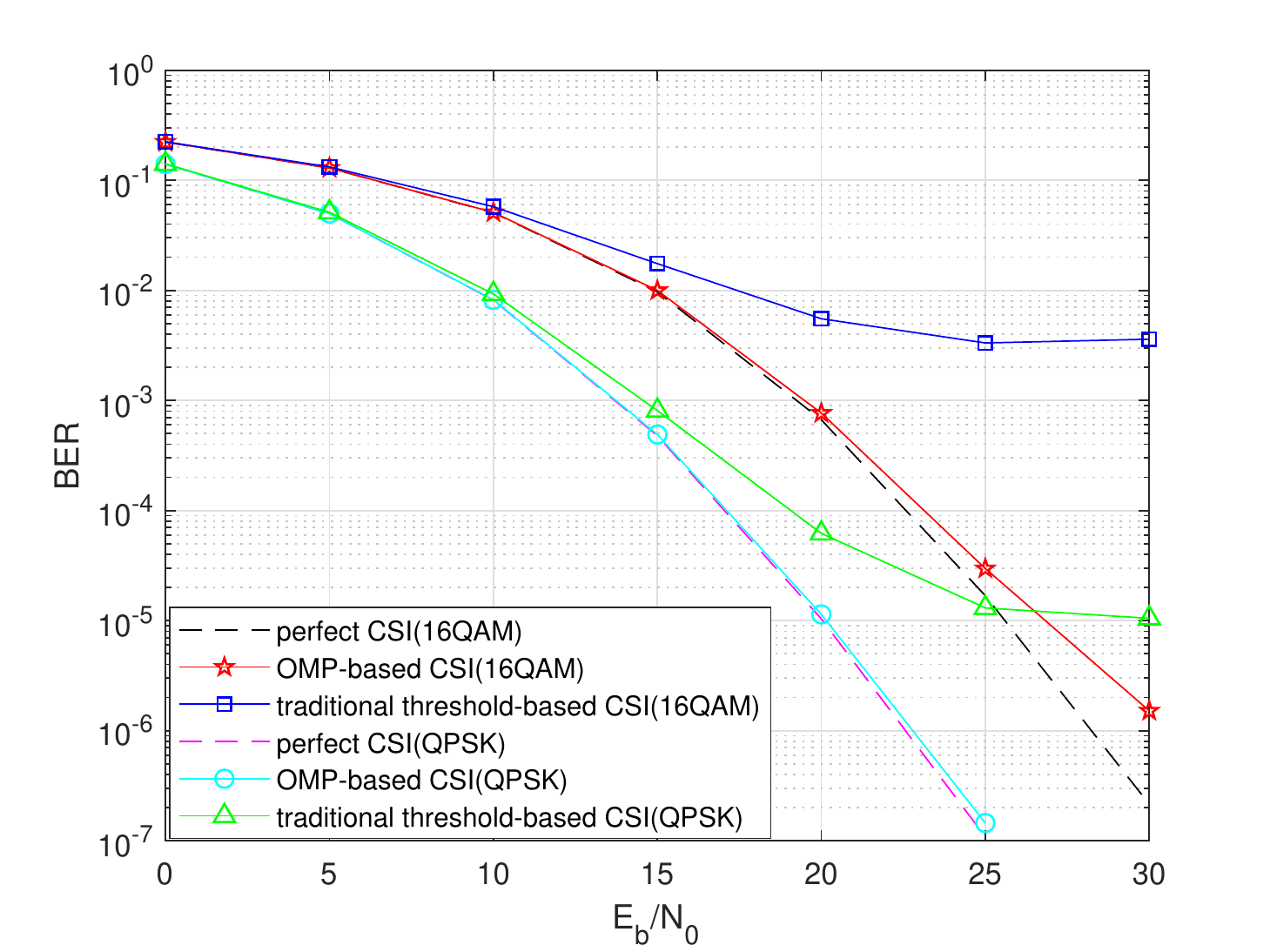}}
		\caption{BER performance under different modulation alphabet caused by different channel estimation scheme against $E_{b}/N_{0}$ with $N=128$, $M=512$, $v=500$ km/h, and $\text{SNR}_{p}=45$dB.}
		\label{Fig_BER_45dB_diffmodulation}	
	\end{figure}
	Finally, we present BER comparison against data SNR $E_{b}/N_{0}$ under different modulation alphabets in Fig. \ref{Fig_BER_45dB_diffmodulation}. $\text{SNR}_{p}$ is set as $45$dB and the velocity is $500$ km/h, which is corresponding to a maximum Doppler frequency of $1.85$kHz. It is obvious that the BER performance employing the OMP-based CSI is nearly the same as perfect CSI when QPSK is selected, however, the BER of threshold-based CSI still has a SNR gap larger than $5$dB when the BER is about $1\times 10^{-5}$. Moreover, our proposed scheme can promote the reliability in high-SNR scenarios by eliminating the error floor caused by inaccurate traditional threshold-based CSI which is about $1\times 10^{-5}$ for QPSK alphabet and $3\times 10^{-3}$ for 16QAM alphabet.\par 
	\section{Conclusion}
	\label{sec:conclusion}
	In this paper, we investigate DSE in OTFS systems for the first time. Specifically, we formulate the delay-Doppler response of the wireless channel and explore the OTFS input-output relationship considering DSE under both ideal (bi-orthogonal) and practical rectangular pulses. Based on the modified input-output relationship considering DSE, the OTFS channel estimation is naturally formulated as a sparse signal recovery problem and an OMP-based channel estimation scheme can be employed directly. Simulation results confirm the necessity to consider DSE and the excellent performance of our estimation scheme taking DSE into consideration. For future research, it is meaningful to consider the pilot optimization to improve the spectral efficiency and efficient off-grid estimation schemes to extract the fractional Doppler.     
	\appendices
	\section{Proof of Theorem \ref{th1_dd_channel}}
	\label{th2_channel_dd_proof}
	Let $H_{i}(t,f)=\beta_{i}
	e^{j2\pi\frac{\nu_{i}}{f_{c}}(f_{c}+f)t} e^{-j2\pi f\tau_{i}}$ denote the time-frequency response of single path and $h_{i}(\tau,\nu)=\iint H_{i}(t,f)e^{-j2\pi(\nu t-\tau f)}dtdf$, we can easily derive that $h(\tau,\nu)=\sum_{i=1}^{N_{P}}h_{i}(\tau,\nu)$, which means we can decouple the problem into finding the delay-Doppler response for single path and directly add them up. Since the case when $\nu_{i}=0$ is trivial, we only offer the proof of the delay-Doppler response for single path with $\nu_{i}\neq 0$, which can be derived as \eqref{h_dd_proof_channel}. Then the proof of Theorem \ref{th1_dd_channel} is completed.
	\begin{equation}
		\scriptsize
		\begin{aligned}
			h_{i}(\tau,\nu)&=\int_{t}\int_{f} H_{i}(t,f)e^{-j2\pi(\nu t-\tau f)}dtdf\\
			&=\int_{t}\int_{f} \beta_{i}
			e^{j2\pi\frac{\nu_{i}}{f_{c}}(f_{c}+f)t}e^{-j2\pi f\tau_{i}}
			e^{-j2\pi(\nu t-\tau f)}dtdf\\
			&=\beta_{i}\int_{t}e^{-j2\pi t(\nu-\nu_{i})}\int_{f}e^{j2\pi f(\tau-\tau_{i}+\frac{\nu_{i}}{f_{c}}t)}dfdt\\
			&=\beta_{i}\int_{t}e^{-j2\pi t(\nu-\nu_{i})}\delta\Big(\frac{\nu_{i}}{f_{c}}t-(\tau_i-\tau)\Big)dt\\
			&=\beta_{i}\Big|\frac{f_{c}}{\nu_{i}}\Big|e^{j2\pi\frac{f_{c}}{\nu_{i}}(\tau-\tau_{i})(\nu-\nu_{i})}\\
		\end{aligned}.
		\label{h_dd_proof_channel}
	\end{equation}
	\begin{figure*}
		\begin{equation}
			\tiny
			\begin{aligned}
				H_{n,m}^{i}[n^{\prime},m^{\prime}]&=\int_{\tau_1^{\prime}}^{\tau_{2}^{\prime}}\int_{\nu_1^{\prime}}^{\nu_{2}^{\prime}}\beta_{i}p_{i}e^{j2\pi p_{i}(\tau-\tau_{i})(\nu-\nu_{i})}e^{j2\pi\nu n^{\prime}T}e^{-j2\pi m\Delta f\tau}e^{j2\pi\nu\tau}d\tau d\nu\\
				&=\beta_{i}p_{i}e^{j2\pi p_{i}\tau_{i}\nu_{i}}\int_{\tau_1^{\prime}}^{\tau_{2}^{\prime}}e^{-j2\pi\tau(m\Delta f+p_{i}\nu_{i})}\int_{\nu_1^{\prime}}^{\nu_{2}^{\prime}}e^{j2\pi\nu\big(\tau+n^{\prime}T-(p_{i}\tau_{i}-p_{i}\tau)\big)}d\nu d\tau\\
				&=\beta_{i}p_{i}e^{j2\pi p_{i}\tau_{i}\nu_{i}}\int_{\tau_1^{\prime}}^{\tau_{2}^{\prime}}e^{-j2\pi\tau(m\Delta f+f_{c})}e^{j2\pi\big((1+p_{i})\tau-(p_{i}\tau_{i}-n^{\prime}T)\big)(m-m^{\prime})\Delta f}\frac{\sin\Big(2\pi f_{\text{max}}\big((1+p_{i})\tau-(p_{i}\tau_{i}-n^{\prime}T)\big)\Big)}{\pi\big((1+p_{i})\tau-(p_{i}\tau_{i}-n^{\prime}T)\big)}d\tau\\
				&=\beta_{i}p_{i}e^{j2\pi p_{i}\tau_{i}\big(\nu_{i}-(m-m^{\prime})\Delta f\big)}\times\int_{\tau_{1}^{\prime}}^{\tau_{2}^{\prime}}\frac{\sin\Big(2\pi f_{\text{max}}\big((1+p_{i})\tau-(p_{i}\tau_{i}-n^{\prime}T)\big)\Big)}{\pi\big((1+p_{i})\tau-(p_{i}\tau_{i}-n^{\prime}T)\big)}e^{-j2\pi\tau\big(f_c+m^{\prime}\Delta f-p_{i}(m-m^{\prime})\Delta f\big)}d\tau\\
			\end{aligned}
			\label{th2_proof_eq}
		\end{equation}
		\hrulefill
		\begin{equation}
			\scriptsize
			\begin{aligned}
				h^{i}[k,l]&=\frac{1}{MN}\sum_{n=0}^{N-1}\sum_{m=0}^{M-1}H^{i}[n,m]e^{-j\pi(\frac{nk}{N}-\frac{ml}{M})}=\frac{1}{MN}\sum_{n=0}^{N-1}\sum_{m=0}^{M-1}\beta_{i}^{\prime}e^{j2\pi (-\tau_{i}m\Delta f+\nu_{i}nT+\frac{mn}{p_{i}})}e^{-j\pi(\frac{nk}{N}-\frac{ml}{M})}\\
				&=\frac{\beta_{i}^{\prime}}{MN}\sum_{n=0}^{N-1}e^{j2\pi n(\nu_{i}T-\frac{k}{N})}\sum_{m=0}^{M-1}e^{-j2\pi m(\tau_{i}\Delta f-\frac{l}{M}-\frac{n}{p_{i}})}\\
				&\overset{(a)}{=}\frac{\beta_{i}^{\prime}}{N}e^{-j\pi(M-1)(\tau_{i}\Delta f-\frac{l}{M})}\sum_{n=0}^{N-1}e^{j2\pi n(\nu_{i}T-\frac{k}{N}+\frac{M-1}{2p_{i}})}\frac{\sin{\pi M(\tau_{i}\Delta f-\frac{l}{M}-\frac{n}{p_{i}})}}{M\sin{\pi(\tau_{i}\Delta f-\frac{l}{M}-\frac{n}{p_{i}})}}\\
				&\overset{(b)}{\approx}\frac{\beta_{i}^{\prime}}{N}e^{-j\pi(M-1)(\tau_{i}\Delta f-\frac{l}{M})}\frac{\sin{\pi M(\tau_{i}\Delta f-\frac{l}{M}-\frac{N-1}{2p_{i}})}}{M\sin{\pi(\tau_{i}\Delta f-\frac{l}{M}-\frac{N-1}{2p_{i}})}}\sum_{n=0}^{N-1}e^{j2\pi n(\nu_{i}T-\frac{k}{N}+\frac{M-1}{2p_{i}})}\\
				&=\beta_{i}^{\prime}e^{-j\pi(M-1)(\frac{l_{i}-l}{M})}e^{j\pi(N-1)(\frac{k_{i}-k}{N})}e^{j\pi\frac{(M-1)(N-1)}{2p_{i}}}\frac{\sin{\pi M(\frac{l_{i}-l}{M}-\frac{N-1}{2p_{i}})}}{M\sin{\pi(\frac{l_{i}-l}{M}-\frac{N-1}{2p_{i}})}}\frac{\sin{\pi N(\frac{k_{i}-k}{N}+\frac{M-1}{2p_{i}})}}{N\sin{\pi (\frac{k_{i-}k}{N}+\frac{M-1}{2p_{i}})}}
			\end{aligned}.
			\label{th3_proof_eq}
		\end{equation}
		\hrulefill
		\begin{equation}
			\tiny
			\label{h_dd_rec_proof_apart}
			\begin{aligned}
				&h_{k,l}^{i}[k^{\prime},l^{\prime}]=\frac{1}{MN}\sum_{n=0}^{N-1}\sum_{m=0}^{M-1}\sum_{n^{\prime}=0}^{N-1}\sum_{m^{\prime}=0}^{M-1}H_{n,m}^{i}[n^{\prime},m^{\prime}]e^{j2\pi(\frac{n^{\prime}k^{\prime}-nk}{N}-\frac{m^{\prime}l^{\prime}-ml}{M})}\\
				&=\frac{1}{MN}\sum_{n,m,n^{\prime},m^{\prime}}\iint h_{i}(\tau,\nu)A_{g_{\text{rx}},g_{\text{tx}}}((n-n^{\prime})T-\tau,(m-m^{\prime})\Delta f-\nu)e^{j2\pi(\nu -m\Delta f)(\tau+n^{\prime}T)}d\tau d\nu e^{j2\pi(\frac{n^{\prime}k^{\prime}-nk}{N}-\frac{m^{\prime}l^{\prime}-ml}{M})}\\
				&=\frac{1}{MN}\sum_{n,m,n^{\prime},m^{\prime}}\iint_{(n-n^{\prime}-1)T}^{(n-n^{\prime})T} \beta_{i}|p_{i}|e^{j2\pi p_{i}(\tau-\tau_{i})(\nu-\nu_{i})}
				\frac{1}{M}\sum_{q\in\mathcal{Q}}e^{j\frac{2\pi q(\nu T+m^{\prime}-m)}{M}}e^{j2\pi(\nu -m\Delta f)(\tau+n^{\prime}T)}d\tau d\nu e^{j2\pi(\frac{n^{\prime}k^{\prime}-nk}{N}-\frac{m^{\prime}l^{\prime}-ml}{M})}\\
				&+\frac{1}{MN}\sum_{n,m,n^{\prime},m^{\prime}}\iint_{(n-n^{\prime})T}^{(n-n^{\prime}+1)T} \beta_{i}|p_{i}|e^{j2\pi p_{i}(\tau-\tau_{i})(\nu-\nu_{i})}\frac{1}{M}\sum_{q\in\mathcal{Q}}e^{j\frac{2\pi q(\nu T+m^{\prime}-m)}{M}}e^{j2\pi(\nu -m\Delta f)(\tau+n^{\prime}T)}d\tau d\nu e^{j2\pi(\frac{n^{\prime}k^{\prime}-nk}{N}-\frac{m^{\prime}l^{\prime}-ml}{M})}\\
			\end{aligned}
		\end{equation}
		\hrulefill
	\end{figure*}
	\section{Proof of Theorem \ref{th2_H_tf_DSE_singlepath}}
	\label{th2_proof_apendix}
	Since $A_{g_{\text{rx}},g_{\text{tx}}}$ holds nonzero only on $[-t_{\text{max}},t_{\text{max}}]\times[-f_{\text{max}},f_{\text{max}}]$, the integral interval in \eqref{H_tf_general} is $[\tau_{1}^{\prime},\tau_{2}^{\prime}]\times[\nu_{1}^{\prime},\nu_{2}^{\prime}]$, where we have
	\begin{equation}
		\tiny
		\begin{cases}
			\tau_{1}^{\prime}=(n-n^{\prime})T-t_{\text{max}}\\
			\tau_{2}^{\prime}=(n-n^{\prime})T+t_{\text{max}}\\
			\nu_{1}^{\prime}=(m-m^{\prime})\Delta f-f_{\text{max}}\\
			\nu_{2}^{\prime}=(m-m^{\prime})\Delta f+f_{\text{max}}
		\end{cases}.
	\end{equation}
	Replacing $h(\tau,\nu)$ in \eqref{H_tf_general} with \eqref{h_DD_wirelesschannel}, the proof of Theorem \ref{th2_H_tf_DSE_singlepath} can be completed by \eqref{th2_proof_eq}.
	\section{Proof of Theorem \ref{th3_h_dd_bi}}
	\label{th3_proof_apendix}
	Since the case $\nu_{i}=0$ is trivial, we only provide the proof for single path when $\nu_{i}\neq 0$.
	$y[k,l]=\sum_{i=1}^{N_{P}}\sum_{k^{\prime}=0}^{N-1}\sum_{l^{\prime}=0}^{M-1}h^{i}[(k-k^{\prime})_{N},(l-l^{\prime})_{M}]x[k^{\prime},l^{\prime}]$ can be obtained by directly employing the SFFT of $Y[n,m]=\sum_{i=1}^{N_{P}}H^{i}[n,m]X[n,m]$, where $h^{i}[k,l]$ can be derived in \eqref{th3_proof_eq} by employing \eqref{H_tf_approx}. Notice that $h^{i}[k.l]$ for $\nu_{i}=0$ can be treated as the limitation when $p_{i}\rightarrow\infty$ consistent with \eqref{th3_proof_eq}, the proof of Theorem \ref{th3_h_dd_bi} is completed by \eqref{th3_proof_eq}.\par
	Notice that we substitute $\frac{\sin{\pi M(\tau_{i}\Delta f-\frac{l}{M}-\frac{n}{p_{i}})}}{M\sin{\pi(\tau_{i}\Delta f-\frac{l}{M}-\frac{n}{p_{i}})}}$ with $\frac{\sin{\pi M(\tau_{i}\Delta f-\frac{l}{M}-\frac{N-1}{2p_{i}})}}{M\sin{\pi(\tau_{i}\Delta f-\frac{l}{M}-\frac{N-1}{2p_{i}})}}$ in \eqref{th3_proof_eq}(b) to provide a closed-form representation, whose approximation precision can be verified by both the NMSE error and BER employing the approximated CSI in Fig. \ref{Fig_NMSE_diffM_perfect} and Fig. \ref{Fig_ber_perfect}. 
	\begin{figure*}
		\begin{equation}
			\tiny
			\label{h_dd_rec_proof_c1}
			\begin{aligned}
				&\frac{1}{MN}\sum_{n,m,n^{\prime},m^{\prime}}\iint_{(n-n^{\prime}-1)T}^{(n-n^{\prime})T} \beta_{i}|p_{i}|e^{j2\pi p_{i}(\tau-\tau_{i})(\nu-\nu_{i})}
				\frac{1}{M}\sum_{q\in\mathcal{Q}}e^{j\frac{2\pi q(\nu T+m^{\prime}-m)}{M}}e^{j2\pi(\nu -m\Delta f)(\tau+n^{\prime}T)}d\tau d\nu e^{j2\pi(\frac{n^{\prime}k^{\prime}-nk}{N}-\frac{m^{\prime}l^{\prime}-ml}{M})}\\
				&=\frac{1}{MN}\sum_{n,m,n^{\prime}}\iint_{(n-n^{\prime}-1)T}^{(n-n^{\prime})T} \beta_{i}|p_{i}|e^{j2\pi p_{i}(\tau-\tau_{i})(\nu-\nu_{i})}e^{j2\pi(\nu -m\Delta f)(\tau+n^{\prime}T)}\frac{1}{M}\sum_{q\in\mathcal{Q}}e^{j2\pi \frac{q(\nu T-m)}{M}}\sum_{m^{\prime}=0}^{M-1}e^{j2\pi \frac{m^{\prime}(q-l^{\prime})}{M}}d\tau d\nu e^{j2\pi(\frac{n^{\prime}k^{\prime}-nk}{N}+\frac{ml}{M})}\\
				&\overset{(a)}{=}\frac{1}{MN}\sum_{n,m,n^{\prime}}\int_{(n-n^{\prime}-1)T}^{(n-n^{\prime})T}\beta_{i}|p_{i}|e^{-j2\pi p_{i}\nu_{i} (\tau-\tau_{i})}e^{-j2\pi\tau m\Delta f}\int e^{j2\pi\nu\big(p_{i}(\tau-\tau_{i})+n^{\prime}T+\tau+\frac{l^{\prime}T}{M}\big)}d\nu d\tau e^{j2\pi(\frac{n^{\prime}k^{\prime}-nk}{N}+\frac{ml-ml^{\prime}}{M})}\\
				&\overset{(b)}{=}\frac{1}{MN}\sum_{n,m,n^{\prime}}\int_{(n-n^{\prime}-1)T}^{(n-n^{\prime})T}\beta_{i}|p_{i}|e^{-j2\pi p_{i}\nu_{i} (\tau-\tau_{i})}e^{-j2\pi\tau m\Delta f}\delta\big((1+p_{i})\tau-(p_{i}\tau_{i}-n^{\prime}T-\frac{l^{\prime}T}{M})\big)d\tau e^{j2\pi(\frac{n^{\prime}k^{\prime}-nk}{N}+\frac{ml-ml^{\prime}}{M})}\\
				&\overset{(c)}{\approx}\frac{\beta_{i}|p_{i}|e^{j2\pi\nu_{i}\tau_{i}}}{MN|1+p_{i}|}e^{j2\pi\nu_{i}\frac{l^{\prime}T}{M}}e^{-j2\pi(\nu_{i}T+\frac{k^{\prime}}{N})}\sum_{n=1}^{N-1}\sum_{m=0}^{M-1}e^{j2\pi\nu_{i}nT}e^{-j2\pi\tau_{i}m\Delta f}e^{j2\pi\frac{m(n-1)}{p_{i}}}e^{j2\pi\frac{n(k^{\prime}-k)}{N}}e^{j2\pi\frac{m(l-l^{\prime})}{M}}e^{j2\pi\frac{ml^{\prime}}{Mp_{i}}}\\
				&\overset{(d)}{\approx}\frac{\beta_{i}^{\prime}}{MN}e^{j2\pi\nu_{i}\frac{l^{\prime}T}{M}}e^{-j2\pi\frac{k_{i}+k^{\prime}}{N}}\sum_{n=1}^{N-1}e^{j2\pi n(\nu_{i}T+\frac{k^{\prime}-k}{N})}\sum_{m=0}^{M-1}e^{-j2\pi m(\tau_{i}\Delta f+\frac{l^{\prime}-l}{M}-\frac{n-1}{p_{i}})}\\
				&=\frac{\beta_{i}^{\prime}}{N}e^{j2\pi\nu_{i}\frac{l^{\prime}T}{M}}e^{-j2\pi\frac{k_{i}+k^{\prime}}{N}}\sum_{n=1}^{N-1}e^{j2\pi n(\nu_{i}T+\frac{k^{\prime}-k}{N})}e^{-j\pi(M-1)(\tau_{i}\Delta f+\frac{l^{\prime}-l}{M}-\frac{n-1}{p_{i}})}\frac{\sin{\pi M(\tau_{i}\Delta f+\frac{l^{\prime}-l}{M}-\frac{n-1}{p_{i}})}}{M\sin{\pi(\tau_{i}\Delta f+\frac{l^{\prime}-l}{M}-\frac{n-1}{p_{i}})}}\\
				&\overset{(e)}{\approx}\frac{\beta_{i}^{\prime}}{N}e^{j2\pi\nu_{i}\frac{l^{\prime}T}{M}}e^{-j2\pi\frac{k_{i}+k^{\prime}}{N}}e^{-j\pi(M-1)(\frac{l_{i}+l^{\prime}-l}{M})}e^{-j\pi\frac{M-1}{p_{i}}}\frac{\sin{\pi M(\frac{l_{i}+l^{\prime}-l}{M}-\frac{N-2}{2p_{i}})}}{M\sin{\pi(\frac{l_{i}+l^{\prime}-l}{M}-\frac{N-2}{2p_{i}})}}\sum_{n=1}^{N-1}e^{j2\pi n(\nu_{i}T+\frac{k^{\prime}-k}{N}+\frac{M-1}{2p_{i}})}\\
				&=\beta_{i}^{\prime}e^{j2\pi\nu_{i}\frac{l^{\prime}T}{M}}e^{-j2\pi\frac{k_{i}+k^{\prime}}{N}}e^{-j\pi(M-1)(\frac{l_{i}+l^{\prime}-l}{M})}e^{-j\pi\frac{M-1}{p_{i}}}\frac{\sin{\pi M(\frac{l_{i}+l^{\prime}-l}{M}-\frac{N-2}{2p_{i}})}}{M\sin{\pi(\frac{l_{i}+l^{\prime}-l}{M}-\frac{N-2}{2p_{i}})}}e^{j\pi N(\frac{k_{i}+k^{\prime}-k}{N}+\frac{M-1}{2p_{i}})}\frac{\sin{\pi(N-1)(\frac{k_{i}+k^{\prime}-k}{N}+\frac{M-1}{2p_{i}})}}{N\sin{\pi(\frac{k_{i}+k^{\prime}-k}{N}+\frac{M-1}{2p_{i}})}}\\
				&=\beta_{i}^{\prime}e^{j2\pi\nu_{i}\frac{l^{\prime}T}{M}}e^{-j2\pi\frac{k_{i}+k^{\prime}}{N}}e^{-j\pi(M-1)(\frac{l_{i}+l^{\prime}-l}{M})}e^{j\pi(k_{i}+k^{\prime}-k)}e^{j\pi\frac{(N-2)(M-1)}{2p_{i}}}\frac{\sin{\pi M(\frac{l_{i}+l^{\prime}-l}{M}-\frac{N-2}{2p_{i}})}}{M\sin{\pi(\frac{l_{i}+l^{\prime}-l}{M}-\frac{N-2}{2p_{i}})}}\frac{\sin{\pi(N-1)(\frac{k_{i}+k^{\prime}-k}{N}+\frac{M-1}{2p_{i}})}}{N\sin{\pi(\frac{k_{i}+k^{\prime}-k}{N}+\frac{M-1}{2p_{i}})}}\\
			\end{aligned}
		\end{equation}
		\hrulefill
		\begin{equation}
			\tiny
			\label{h_dd_rec_proof_c2}
			\begin{aligned}
				&\frac{1}{MN}\sum_{n,m,n^{\prime},m^{\prime}}\iint_{(n-n^{\prime})T}^{(n-n^{\prime}+1)T} \beta_{i}|p_{i}|e^{j2\pi p_{i}(\tau-\tau_{i})(\nu-\nu_{i})}\frac{1}{M}\sum_{q\in\mathcal{Q}}e^{j\frac{2\pi q(\nu T+m^{\prime}-m)}{M}}e^{j2\pi(\nu -m\Delta f)(\tau+n^{\prime}T)}d\tau d\nu e^{j2\pi(\frac{n^{\prime}k^{\prime}-nk}{N}-\frac{m^{\prime}l^{\prime}-ml}{M})}\\
				&=\frac{1}{MN}\sum_{n,m,n^{\prime}}\iint_{(n-n^{\prime})T}^{(n-n^{\prime}+1)T}\beta_{i}|p_{i}|e^{j2\pi p_{i}(\tau-\tau_{i})(\nu-\nu_{i})}e^{j2\pi(\nu -m\Delta f)(\tau+n^{\prime}T)}\frac{1}{M}\sum_{q\in\mathcal{Q}}e^{j\frac{2\pi q(\nu T-m)}{M}}\sum_{m^{\prime}=0}^{M-1}e^{j2\pi\frac{m^{\prime}(q-l^{\prime})}{M}}d\tau d\nu e^{j2\pi(\frac{n^{\prime}k^{\prime}-nk}{N}+\frac{ml}{M})}\\
				&\overset{(a)}{=}\frac{1}{MN}\sum_{n,m,n^{\prime}}\int_{(n-n^{\prime})T}^{(n-n^{\prime}+1)T} \beta_{i}|p_{i}|e^{-j2\pi p_{i}\nu_{i} (\tau-\tau_{i})}e^{-j2\pi\tau m\Delta f}\int e^{j2\pi\nu\big(p_{i}(\tau-\tau_{i})+n^{\prime}T+\tau+\frac{l^{\prime}T}{M}\big)}d\nu d\tau e^{j2\pi(\frac{n^{\prime}k^{\prime}-nk}{N}+\frac{ml-ml^{\prime}}{M})}\\
				&\overset{(b)}{=}\frac{1}{MN}\sum_{n,m,n^{\prime}}\int_{(n-n^{\prime}-1)T}^{(n-n^{\prime})T}\beta_{i}|p_{i}|e^{-j2\pi p_{i}\nu_{i} (\tau-\tau_{i})}e^{-j2\pi\tau m\Delta f}\delta\big((1+p_{i})\tau-(p_{i}\tau_{i}-n^{\prime}T-\frac{l^{\prime}T}{M})\big)d\tau e^{j2\pi(\frac{n^{\prime}k^{\prime}-nk}{N}+\frac{ml-ml^{\prime}}{M})}\\
				&\overset{(c)}{\approx}\frac{\beta_{i}|p_{i}|e^{j2\pi\nu_{i}\tau_{i}}}{MN|1+p_{i}|}e^{j2\pi\nu_{i}\frac{l^{\prime}T}{M}}\sum_{n=0}^{N-1}\sum_{m=0}^{M-1}e^{j2\pi\nu_{i}nT}e^{-j2\pi\tau_{i}m\Delta f}e^{j2\pi\frac{mn}{p_{i}}}e^{j2\pi\frac{n(k^{\prime}-k)}{N}}e^{j2\pi\frac{m(l-l^{\prime})}{M}}e^{j2\pi\frac{ml^{\prime}}{Mp_{i}}}\\
				&\overset{(d)}{\approx}\frac{\beta_{i}^{\prime}}{MN}e^{j2\pi\nu_{i}\frac{l^{\prime}T}{M}}\sum_{n=0}^{N-1}e^{j2\pi n(\nu_{i}T+\frac{k^{\prime}-k}{N})}\sum_{m=0}^{M-1}e^{-j2\pi m(\tau_{i}\Delta f+\frac{l^{\prime}-l}{M}-\frac{n}{p_{i}})}\\
				&=\frac{\beta_{i}^{\prime}}{N}e^{j2\pi\nu_{i}\frac{l^{\prime}T}{M}}e^{-j\pi(M-1)(\frac{l_{i}+l^{\prime}-l}{M})}\sum_{n=0}^{N-1}e^{j2\pi n(\nu_{i}T+\frac{k^{\prime}-k}{N}+\frac{M-1}{2p_{i}})}\frac{\sin{\pi M(\tau_{i}\Delta f+\frac{l^{\prime}-l}{M}-\frac{n}{p_{i}})}}{M\sin{\pi(\tau_{i}\Delta f+\frac{l^{\prime}-l}{M}-\frac{n}{p_{i}})}}\\
				&\overset{(e)}{\approx}\frac{\beta_{i}^{\prime}}{N}e^{j2\pi\nu_{i}\frac{l^{\prime}T}{M}}e^{-j\pi(M-1)(\frac{l_{i}+l^{\prime}-l}{M})}\frac{\sin{\pi M(\tau_{i}\Delta f+\frac{l^{\prime}-l}{M}-\frac{N-1}{2p_{i}})}}{M\sin{\pi(\tau_{i}\Delta f+\frac{l^{\prime}-l}{M}-\frac{N-1}{2p_{i}})}}\sum_{n=0}^{N-1}e^{j2\pi n(\nu_{i}T+\frac{k^{\prime}-k}{N}+\frac{M-1}{2p_{i}})}\\
				&=\beta_{i}^{\prime}e^{j2\pi\nu_{i}\frac{l^{\prime}T}{M}}e^{-j\pi(M-1)(\frac{l_{i}+l^{\prime}-l}{M})}e^{j\pi(N-1)\frac{k_{i}+k^{\prime}-k}{N}}e^{j\pi\frac{(N-1)(M-1)}{2p_{i}}}\frac{\sin{\pi M(\frac{l_{i}+l^{\prime}-l}{M}-\frac{N-1}{2p_{i}})}}{M\sin{\pi(\frac{l_{i}+l^{\prime}-l}{M}-\frac{N-1}{2p_{i}})}}\frac{\sin{\pi N(\frac{k_{i}+k^{\prime}-k}{N}+\frac{M-1}{2p_{i}})}}{N\sin{\pi(\frac{k_{i}+k^{\prime}-k}{N}+\frac{M-1}{2p_{i}})}}
			\end{aligned}
		\end{equation}
	\end{figure*}
	\section{Proof of Theorem \ref{th4_h_dd_rec}}
	\label{th4_proof_apendix}
	At first, the cross-ambiguity approximation in \eqref{cross_appro_sum_rec} and the LTV channel response in \eqref{h_DD_wirelesschannel} is substituted in generalized representation in \eqref{h_dd_rec_proof_apart}, where we find the $h^{i}_{k,l}[k^{\prime},l^{\prime}]$ can be divided into two parts as \eqref{h_dd_rec_proof_c1} and \eqref{h_dd_rec_proof_c2}.\par
	For \eqref{h_dd_rec_proof_c1}, (a) is obtained since $\sum_{m^{\prime}=0}^{M-1}e^{j2\pi\frac{m^{\prime}(q-l^{\prime})}{M}}$ is non-zero only when $q=l^{\prime}$, which indicates $l^{\prime}\in\mathcal{Q}$ and derives (b). Since $\delta(x)$ is non-zero only when $x=0$, we have 
	\begin{equation}
		\label{c1_range_nn'_pi+}
		\scriptsize
		\begin{cases}
			(1+p_{i})\big((n-n^{\prime})T-\frac{l^{\prime}T}{M}\big)-(p_{i}\tau_{i}-n^{\prime}T-\frac{l^{\prime}T}{M})<0\\
			(1+p_{i})((n-n^{\prime})T)-(p_{i}\tau_{i}-n^{\prime}T-\frac{l^{\prime}T}{M})>0
		\end{cases}
	\end{equation}
	for $p_{i}>0$ and
	\begin{equation}
		\scriptsize
		\label{c1_range_nn'_pi-}
		\begin{cases}
			(1+p_{i})\big((n-n^{\prime})T-\frac{l^{\prime}T}{M}\big)-(p_{i}\tau_{i}-n^{\prime}T-\frac{l^{\prime}T}{M})>0\\
			(1+p_{i})((n-n^{\prime})T)-(p_{i}\tau_{i}-n^{\prime}T-\frac{l^{\prime}T}{M})<0
		\end{cases}
	\end{equation}
	for $p_{i}<0$, where the lower limit $(n-n^{\prime}-1)T$ is replaced with $(n-n^{\prime})T-\frac{l^{\prime}T}{M}$ to satisfy $l^{\prime}\in\mathcal{Q}$. Combining \eqref{c1_range_nn'_pi+} with \eqref{c1_range_nn'_pi-}, the range of $(n-n^{\prime})T$ can be derived as
	\begin{equation}
		\scriptsize
		\tau_{i}-\frac{nT+\frac{l^{\prime}T}{M}}{p_{i}}<(n-n^{\prime})T<\tau_{i}+\frac{l^{\prime}T}{M}-\frac{nT}{p_{i}}.
		\label{range_nn'_initial_rec_1}
	\end{equation}
	\eqref{range_nn'_initial_rec_1} can be proceeded further by employing $|p_{i}|>MN$ and $\frac{T}{M}\leq\tau_{i}<T$, where we have
	\begin{equation}
		\scriptsize
		\tau_{i}-\frac{nT+\frac{l^{\prime}T}{M}}{p_{i}}\geq\tau_{i}-\frac{(N-1)T+T}{|p_{i}|}>\tau_{i}-\frac{T}{M}\geq 0
		\label{range_nn'_rec_c1_left}
	\end{equation}
	and
	\begin{equation}
		\scriptsize
		\tau_{i}+\frac{l^{\prime}T}{M}-\frac{nT}{p_{i}}<T+\frac{(M-1)T}{M}+\frac{(N-1)T}{MN}<2T.
		\label{range_nn'_rec_c1_right}
	\end{equation}
	As a result, $n-n^{\prime}=1$ and $n\geq 1$ will be attained, which helps derive the range of $n$ for non-zero integral as
	\begin{equation}
		\scriptsize
		\frac{nT}{p_{i}}<\frac{l^{\prime}T}{M}+\tau_{i}-T=\frac{l^{\prime}+l_{i}-M}{M}T.
		\label{nrange_c1}
	\end{equation}
	Since we have assumed integer delay for wideband OTFS systems and $\frac{nT}{|p_{i}|}\in(0,\frac{T}{M})$, \eqref{nrange_c1} holds either true or false for $\forall 1\leq n\leq N-1$, which provides
	\begin{equation}
		\scriptsize
		\mathcal{L}_{ISI}^{i}=
		\begin{cases}
			\{l^{\prime}\in \mathbb{N}:M-l_{i}+1\leq l^{\prime}\leq M-1\},&p_{i}>0\\
			\{l^{\prime}\in \mathbb{N}:M-l_{i}\leq l^{\prime}\leq M-1\},&p_{i}<0\\
		\end{cases}
	\end{equation}\par
	When $l^{\prime}\in\mathcal{L}_{ISI}^{i}$, \eqref{h_dd_rec_proof_c1} is effective since \eqref{nrange_c1} holds true for $\forall 1\leq n\leq N-1$. If $l^{\prime}\notin\mathcal{L}_{ISI}^{i}$, this part becomes $0$ since \eqref{nrange_c1} holds false for $\forall 1\leq n\leq N-1$. Then (c) is obtained by employing $n-n^{\prime}=1$ and the property of delta function. The approximation in (c) and (d) is based on $|p_{i}|\gg 1$, which derives $1+p_{i}\approx p_{i}$ and $1+\frac{1}{p_{i}}\approx 1$. The approximation in (e) is similar to \eqref{th3_proof_eq}(b), where $\frac{n-1}{p_{i}}$ in the phase of discrete sinc function is substituted by the median $\frac{N-2}{2p_{i}}$.\par 
	The analysis for \eqref{h_dd_rec_proof_c2} is similar to \eqref{h_dd_rec_proof_c1}. (a) is obtained since $\sum_{m^{\prime}=0}^{M-1}e^{j2\pi\frac{m^{\prime}(q-l^{\prime})}{M}}$ is non-zero only when $q=l^{\prime}$, which indicates $l^{\prime}\in\mathcal{Q}$ and derives (b). Since $\delta(x)$ is non-zero only when $x=0$, we have 
	\begin{equation}
		\label{c2_range_nn'_pi+}
		\scriptsize
		\begin{cases}
			(1+p_{i})((n-n^{\prime})T)-p_{i}\tau_{i}+n^{\prime}T+\frac{l^{\prime}T}{M}<0\\
			(1+p_{i})\big((n-n^{\prime}+1)T-\frac{l^{\prime}T}{M}\big)-p_{i}\tau_{i}+n^{\prime}T+\frac{l^{\prime}T}{M}>0
		\end{cases}
	\end{equation}
	for $p_{i}>0$ and
	\begin{equation}
		\label{c2_range_nn'_pi-}
		\scriptsize
		\begin{cases}
			(1+p_{i})((n-n^{\prime})T)-p_{i}\tau_{i}+n^{\prime}T+\frac{l^{\prime}T}{M}>0\\
			(1+p_{i})\big((n-n^{\prime}+1)T-\frac{l^{\prime}T}{M}\big)-p_{i}\tau_{i}+n^{\prime}T+\frac{l^{\prime}T}{M}<0
		\end{cases}
	\end{equation}
	for $p_{i}<0$, where the upper limit $(n-n^{\prime}+1)T$ is replaced with $(n-n^{\prime}+1)T-\frac{l^{\prime}T}{M}$ to satisfy $l^{\prime}\in\mathcal{Q}$. Combining \eqref{c2_range_nn'_pi+} with \eqref{c2_range_nn'_pi-}, the range of $(n-n^{\prime})T$ can be derived as
	\begin{equation}
		\scriptsize
		-T+\tau_{i}+\frac{l^{\prime}T}{M}-\frac{(n+1)T}{p_{i}}<(n-n^{\prime})T<\tau_{i}-\frac{nT+\frac{l^{\prime}T}{M}}{p_{i}}.
		\label{range_nn'_initial_rec_0}
	\end{equation}
	\eqref{range_nn'_initial_rec_0} can be dug further by employing $|p_{i}|>MN$ and $\frac{T}{M}\leq\tau_{i}<T$, where we have
	\begin{equation}
		\scriptsize
		-T+\tau_{i}+\frac{l^{\prime}T}{M}-\frac{(n+1)T}{p_{i}}>-T+(\tau_{i}-\frac{NT}{|p_{i}|})>-T.
		\label{range_nn'_rec_c2_left}
	\end{equation} 
	Combining the result of \eqref{range_nn'_rec_c2_left} with $\tau_{i}+\frac{T}{M}<T$, it is obvious that this part is possibly non-zero only $n=n^{'}$, which helps derive the range of $n$ for non-zero integral as
	\begin{equation}
		\scriptsize
		\frac{(n+1)T}{p_{i}}>\frac{l^{\prime}T}{M}+\tau_{i}-T=\frac{l^{\prime}+l_{i}-M}{M}T.
		\label{nrange_c2}
	\end{equation} 
	Since we have assumed integer delay for wideband OTFS systems and $\frac{(n+1)T}{|p_{i}|}\in(0,\frac{T}{M})$, \eqref{nrange_c2} holds either true or false for $\forall 1\leq n\leq N-1$, which offers
	\begin{equation}
		\scriptsize
		\mathcal{L}_{ICI}^{i}=
		\begin{cases}
			\{l^{\prime}\in \mathbb{N}:0\leq l^{\prime}\leq M-l_{i}\},&p_{i}>0\\
			\{l^{\prime}\in \mathbb{N}:0\leq l^{\prime}\leq M-l_{i}-1\},&p_{i}<0\\
		\end{cases}
	\end{equation}\par
	When $l^{\prime}\in\mathcal{L}_{ICI}^{i}$, \eqref{h_dd_rec_proof_c2} is effective since \eqref{nrange_c2} holds true for $\forall 1\leq n\leq N-1$. If $l^{\prime}\notin\mathcal{L}_{ICI}^{i}$, this part becomes $0$ since \eqref{nrange_c2} holds false for $\forall 1\leq n\leq N-1$. Then (c) is obtained by employing $n=n^{\prime}$ and the property of delta function. The approximation in (c) and (d) is based on $|p_{i}|\gg 1$, which derives $1+p_{i}\approx p_{i}$ and $1+\frac{1}{p_{i}}\approx 1$. The approximation in (e) is similar to \eqref{th3_proof_eq}(b), where $\frac{n}{p_{i}}$ in the phase of discrete sinc function is substituted by the median $\frac{N-1}{2p_{i}}$.\par 
	The proof of Theorem \ref{th4_h_dd_rec} is completed by combining the deduction of \eqref{h_dd_rec_proof_c1} and \eqref{h_dd_rec_proof_c2} with the extent analysis before.
	\color{black}
	\bibliographystyle{IEEEtran}
	\bibliography{ref-sum}
\end{spacing}	
\end{document}